\documentstyle[12pt,cite,epsfig,psfrag]{article}

\def\be{\begin{equation}}
\def\ee{\end{equation}}
\def\bea{\begin{eqnarray}}
\def\eea{\end{eqnarray}}
\def\pa{\partial}

\newcommand{\beqal}{\begin{eqnarray}\label}
\newcommand{\beqa}{\begin{eqnarray}}
\newcommand{\eeqa}{\end{eqnarray}}

\newcommand {\Tr}{\mbox{Tr}}

\newcommand {\non}{\\\nonumber}
\newcommand {\f}{\frac}

\begin{document}
\baselineskip=0.6cm
\begin{titlepage}
\begin{center}
%\hfill hep-th/yymmnnn\\
%\hfill IP/BBSR/2007-??\\
\vskip .2in

{\Large \bf Phase transitions in higher derivative gravity and gauge theory: R-charged
black holes}
\vskip .5in

{\bf Tanay K. Dey$^*$\footnote{e-mail: tanay@iopb.res.in}, Sudipta
Mukherji$^*$\footnote{e-mail: mukherji@iopb.res.in},
Subir Mukhopadhyay$^{*+}$\footnote{e-mail: subir@iopb.res.in} and
Swarnendu Sarkar$^*$\footnote{e-mail: swarnen@iopb.res.in}\\
\vskip .1in
{\em $^*$Institute of Physics,\\
Bhubaneswar 751 005, INDIA}
\vskip .1in
{\em $^+$Institute for Studies in Theoretical Physics and Mathematics
(IPM),\\
P.O.Box 19395-5531, Tehran, IRAN}}

\end{center}

\begin{center} {\bf ABSTRACT}
\end{center}
\begin{quotation}\noindent
\baselineskip 15pt

This is a continuation of our earlier work where we constructed a
phenomenologically motivated effective action of the boundary gauge
theory at finite temperature and finite gauge coupling on $S^3
\times S^1$. In this paper, we argue that this effective action
qualitatively reproduces the gauge theory representing
various bulk phases of R-charged black hole with Gauss-Bonnet
correction. We analyze the system both in canonical and grand
canonical ensemble.

\end{quotation}
\vskip 2in
June 2007\\
\end{titlepage}
\vfill
\eject

\section{Introduction and Summary}
%%%%%%

In a recent paper \cite{dmms}, we initiated a study of the phase
structure of gravity in anti-de Sitter space in presence of higher
derivative corrections. Our motivation stemmed from the fact that
due to AdS/CFT correspondence \cite{Maldacena:1997re,Witten:1998qj}
a quantum theory of gravity provides natural arena for addressing
issues in gauge theory and vice versa. Though absence of a
formulation of string theory on AdS space and lack of adequate
techniques to study strong coupling regimes of gauge theories make a
quantitative comparison difficult, nevertheless, encouraging
agreements among qualitative features have been gathered over the
last few years. One of the remarkable steps\cite{wittenone} in this
realm is the identification of the crossover from the thermal AdS
phase to black hole phase, namely the Hawking-Page
transition\cite{H-P}, on the gravity side with large $N$ deconfinement
transition on the gauge theory side \footnote{In many papers, it was
argued that this large $N$ transition might even occur at zero
coupling \cite{bs, ammpr} and it turned out that for zero coupling,
the transition appeared exactly at the Hagedron temperature of the
low temperature thermal AdS phase. As the coupling is increased the
Hagedorn and the deconfinement transitions separate out. The
Hagedorn transition occurs at a higher temperature than that of the
deconfinement transition. However it is also believed that this
Hagedorn transition at weak coupling manifests as the deconfinement
or Hawking-Page transition at strong coupling. For a related work
see \cite{RS}. Non-perturbative effects near Hagedorn transition was
studied in \cite{hl} (see also \cite{Liu}}. Based on this
identification, recently, a phenomenological matrix model was
proposed\cite{aglw}, which belongs to the same universality class of
${\cal N}$=4 supersymmetric $SU(N)$ gauge theory on $S^3$ at the limit
of infinite 't Hooft coupling ($\lambda$). This model, which we will
call $(a,b)$-model in the sequel, was characterized by two
coefficients $a$ and $b$ which depend on temperature $T$ (and
$\lambda$). It correctly reproduced the qualitative features of the
phase structure of the dual theory on the gravity side. This model
was also used to study the transition between AdS soliton and black
hole\cite{BE}

Following this line of works, in \cite{dmms} we considered the
response of the Hawking-Page transition and the associated
thermodynamic phase structure to higher derivative terms in the
gravity theory, which appears as $\alpha^\prime$ corrections of the
underlying string theory. While a study with general class of higher
derivative terms would be desirable, in \cite{dmms}, we restricted
ourselves to the Gauss-Bonnet (GB) terms only. The reason for this is that
the GB terms can capture non-trivial first order effects of $\alpha^\prime$
correction while admitting explicit black hole
solutions\cite{BD,RM,Nojiri:2001aj,RCAI,CNO,CN,Neupane,TM} \footnote{GB term
is expected to occur only in Heterotic or K3 compactification of
type IIA theory but not in type II theories with maximally
supersymmetric compactification. The lowest correction in type IIB
theory on $AdS_5$ is of order $\alpha^{'3}$. The thermodynamic
phases of the perturbative supergravity solutions as well as their
boundary duals have been studied by various authors
\cite{GUB,GL,KL,CK}. However the qualitative phase structures in
this case is quite similar to that of gravity with GB
($\alpha^\prime$) correction. }. It turned out that the phase
structure in GB theory depends crucially on the GB coupling. For
sufficiently strong coupling there is a single big black hole phase
which serves as a local minimum below a critical temperature. As the
coupling reduces below a critical value, two additional black holes
of intermediate and small sizes appear, of which, the former one has
negative specific heat while the small hole has positive specific
heat. The latter one is stable and corresponds to a local minimum up
to a critical temperature beyond which it ceases to exist. From the
associated phase diagram, we find that the phase structure is
similar to that of a Van der Wall's gas. Along with the standard HP transition
between the big black hole and thermal AdS, we identified one more
phase transition where crossover of the energies of small black hole
and big black hole occurs.

A similar study of the various phases in the dual theory is also
carried out in \cite{dmms}. Usually, to describe gauge theory on
$S^3\times S^1$ at zero coupling one writes down an effective action
with vev of Polyakov loop on $S^1$ as the only light degree of
freedom. Since it is hard to compute such an action at finite YM
coupling, one relies on phenomenological model\cite{aglw}, that
belongs to the same universality class of the gauge theory as
described in\cite{aglw} . Appealing to the universal nature of this
model near the critical temperature, we analyzed in \cite{dmms} the
$\frac{1}{\lambda}$ dependence of the coefficients $(a,b)$ by
mapping the GB corrections to the gauge theory. We also
suggest\cite{dmms} a modified matrix model which, with appropriate
restriction on the coefficients of the model, qualitatively captures
the bulk properties with $\alpha^\prime$ corrections. This is an
extension of $(a,b)$-model and includes higher power of density of
eigenvalues of the vev of Polyakov loop indicating, perhaps, that
other operators in matrix model become relevant.

In this paper, we extend our study of the effects of higher
curvature corrections to charged sector in the AdS/CFT set up with
an aim to identify some universal features of the boundary matrix
models at strong coupling. The charged sector of the GB action
contains a Maxwell term besides the GB corrections. Maxwell term
typically comes in type IIB theory on $AdS_5 \times S^5$ from
angular momentum along $S^5$ direction, or in other words, from the
$SO(6)$ gauge symmetry arising from the group of isometries of
$S^5$. We focus our attention to those black hole configurations
which have equal $U(1)$ charges for all the three commuting $U(1)$
subgroups of $SO(6)$. These black holes and their phase structures
were considered in \cite{cejmo, cejmt} (see also \cite{ttorii}). We
study the changes of phase structures due to GB corrections in
canonical and grand canonical ensembles. On the gauge theory side
\footnote{We would like to emphasise that the inclusion of GB correction takes us away from type IIB framework. However we will assume that a version of 
gauge/gravity will continue to hold.},
in order to describe the charged sector, we use the same model as
in\cite{dmms} except we allow the coefficients of the model to
depend on appropriate parameters of the ensemble as well along with
the temperature ($T$) and the t'Hooft coupling ($\lambda$). In the
following we summarize our results along with some already known
facts about phase diagram in absence of GB coupling.
\\

\noindent {\bf Grand canonical Ensemble (Fixed Potential):}

\noindent In the grand canonical ensemble, the black hole is allowed to
emit and absorb charged particles keeping the potential fixed till
the thermal equilibrium is reached, which, in this case, is governed
by a fixed chemical potential. Here the phase diagram is
characterized by the chemical potential $\Phi$ and the GB coupling which we call
$\bar{\alpha}$ in the paper.

$(\Phi \ne 0,\bar{\alpha} = 0)$ : On the gravity side, the phase
diagrams have been analyzed in \cite{cejmo, cejmt}. If the potential
$\Phi$ is below a critical value, various phases are similar to that
of AdS-Schwarzschild black hole while for $\Phi$ large enough, the
black hole free energy becomes negative compared to that of the
thermal AdS at any fixed temperature. On the gauge theory side, at
zero and small $\lambda$, the phase structure was analysed in
\cite{yy,TM}, while for large $\lambda$, a phenomenologically motivated
matrix model can be constructed and we will have occasion to
elaborate on it at a later stage of this paper.

$(\Phi \ne 0,\bar{\alpha} \ne 0)$ : This case is studied in section
(2.1) where we find the critical value of $\Phi$ depends on
$\alpha$. For small $\Phi$, there are three different black hole
phases; one of them being unstable. Identifying the rest two as a
small and a big black hole, we find that there is a first order
phase transition from the small to the big black hole. However, once
thermal AdS is included in the phase diagram, we find both the small
and big black hole phases are metastable at low temperature and big
black hole becomes stable only at high temperature. In order to
clearly illustrate the various phases in this range of $\Phi$ we
construct a Landau function with black hole horizon radius as the
order parameter. When $\Phi$ is above the critical value, phase
diagram shows a single black hole phase which is stable beyond
certain temperature while a crossover from black hole phase to
thermal AdS occurs for temperature lower than that. If we increase
$\Phi$ even further, the black hole phase is always found to be
stable. All these phases can be summarized in a ($\Phi^2 -
\bar\alpha$) diagram; see Figure \ref{alphaqphia}. We also note that, in all
the above cases, wherever there is Hawking-Page transition from AdS
to the black hole phase, the transition temperature is found to
decrease with $\bar \alpha$.

We study the grand canonical ensemble of the dual theory in section
(3.3) and (3.4). Here the parameters of the matrix model depend on
the chemical potential ($\mu$). We find for the chemical potential
less than the critical value the analysis is similar to that of
\cite{dmms}. As in \cite{dmms}, the matrix model has an extra saddle
point that has no analogue in supergravity. We interpret this saddle
point with some phase in string theory\footnote{ This could be
related to some state in string theory. At this point it may be
mentioned that appearance of string states in boundary theory also
occurred in \cite{aglw,Alvarez-Gaume:2006jg} which corresponds to
the Gross-Witten transition \cite{Gross:1980he,Wadia:1980cp} and
which was identified with a crossover from supergravity black hole
solution to string state \cite{Horowitz:1996nw}.} Beyond the
critical potential we encounter different possible situations
depending on the position (expectation value of the Polyakov loop)
of the extra saddle point.\\

\noindent {\bf Canonical ensemble (Fixed Charge):}

\noindent In the canonical ensemble, the black hole is allowed to emit
and absorb radiation, keeping the charge fixed till the thermal
equilibrium is reached and the phase diagrams are characterized by the charge of the black hole, $q$ and the GB coupling
$\bar{\alpha}$.

$(q \ne 0,\bar{\alpha}=0)$ : The phase structure is discussed in
\cite{cejmo, cejmt} in great detail. There exists a critical charge
$q_c$ above which, at all temperature, only one black hole phase
exists. Below $q_c$, there can be at most three black hole phases.
We call them  small, intermediate and large. While the intermediate
one is unstable, the small and big black holes are stable. It was
also noted that thermal AdS is not an admissible phase. When we
increase the temperature, there is a crossover from a small black
hole to a large black hole phase via a first order phase transition.

$(q \ne 0,\bar{\alpha} \ne 0)$ : This part of the analysis is given
in section 2.2, where, we find the phase structure depends on two
parameters $q$ and $\bar\alpha$. In particular, in ($q^2 - \bar
\alpha$) plane, we identify two distinct regions (see 
Figure \ref{alphaqphib}) where region I consists of three black hole
phases, while in region II, only one black hole phase exists.
Thermal AdS continues to be non-admissible phase. As before, there
is a transition from  small to big black hole at a critical
temperature. This temperature decreases as we increase $\bar\alpha$.

Study of the canonical ensemble of the dual theory is discussed in
section (3.5). Since the explicit dependence of the coefficients of
the matrix model on the chemical potential is very hard to determine
at strong coupling, we assume the the zero coupling result is valid
there or at least the universality class of the theory does not
change once we tune up the gauge coupling. We find that this
dependence is consistent with one of the possible scenarios. For
this scenario, we write down the matrix model for the fixed charge
and find that this model correctly reproduces the phases of the
black holes with fixed charge.

The paper is structured as follows. We begin with the thermodynamics
of charged sector in presence of GB coupling in section 2.
Subsections (2.1) and (2.2) are devoted to discussion of canonical
and grand canonical ensembles. The dual theory is discussed in
section 3. We begin by reviewing zero charge sector of the dual
theory. A short account of the zero and weak 't Hooft coupling
results is given in subsection (3.1) and in subsection (3.2) we
briefly mention the comparison of phase structures between gravity
and matrix model. The rest of the section 3 is devoted to our study
of the charged sector. The grand canonical ensemble of the dual
theory is considered in subsection (3.3) and (3.4) while the
canonical ensemble is discussed in subsection (3.5).
%%%%%%
%\newpage
%%%%%%%%%%%

\section{Gauss-Bonnet black hole with electric charge}\label{bulk}

We start with $n+1$-dimensional $(n \ge 4)$ action
\begin{equation}
I = \int d^{n+1}x {\sqrt{-g_{n+1}}}\Big[ {R\over\kappa_{n+1}} - 2 \Lambda
+ \alpha (R^2 - 4 R_{ab}R^{ab} + R_{abcd}R^{abcd}) - {F^2\over \kappa_{n+1}}\Big].
\label{act}
\end{equation}
Here, $\alpha$ is the GB coupling. As the higher derivative corrections are expected
to appear from the $\alpha^\prime$ corrections in underlying string theory, we will
often refer the GB term as $\alpha^\prime$ correction in this paper.
This action possesses black hole solutions which we call charged
GB black holes \cite{ttorii}. These solutions have the form
\begin{equation}
ds^2 = - V(r) dt^2 + {dr^2\over{V(r)}} + r^2 d\Omega^2_{n-1},
\label{gbmetric}
\end{equation}
where $V(r)$ is given by
\begin{equation}
V(r) = 1 + {r^2\over{2\hat\alpha}} - {r^2\over{2\hat\alpha}}\Big[
1 - {4 \hat\alpha\over l^2} + {4\hat\alpha m\over{r^{n}}} -
{4 \hat\alpha q^2\over{r^{2(n-1)}}}\Big]^{1\over 2}.
\label{cgbsol}
\end{equation}
In the above, $\hat \alpha = (n-2)(n-3) \alpha \kappa_{n+1}$, while $l$ is related to the
cosmological constant as $l^2 = -n(n-1)/(2\kappa_{n+1}\Lambda)$. The parameter $m$
is related to the ADM mass of the hole, $M$ as
\begin{equation}
M = {(n-1) \omega_{n-1} m\over {\kappa_{n+1}}},
\end{equation}
where $\omega_{n-1}$ is the volume of the unit $(n-1)$-sphere. The parameter $q$ gives
the charge
\begin{equation}
Q = {{2 \sqrt{2 (n-1)(n-2)}} \omega_{n-1} q \over \kappa_{n+1}},
\end{equation}
of the electric gauge potential
\begin{equation}
A_t = - {\sqrt {n -1\over {2 (n-2)}}} {q \over {r^{n-2}}} + \Phi,
\label{gaugepot}
\end{equation}
where $\Phi$ is a constant which we will fix below. Denoting $r_+$ as the
largest real positive root of $V(r)$, we find that the metric (\ref{cgbsol})
describes a black hole with non-singular horizon if
\begin{equation}
\Big({n\over{n-2}}\Big) r_+^{2n -2} + l^2 r_+^{2n-4} \ge q^2 l^2.
\end{equation}

Finally, we shall choose the gauge potential $A_t$ to vanish at the
horizon. This fixes $\Phi$ to be
\begin{equation}
\Phi = {\sqrt {n -1\over {2 (n-2)}}} {q \over {{r_+}^{n-2}}}.
\end{equation}
This quantity is the electrostatic potential between the
horizon and infinity. Asymptotically, the metric (\ref{cgbsol}) goes to AdS space, as
in
this limit,
\begin{equation}
V(r) = 1 + \Big[{1\over {2 \hat\alpha}} - {1\over{2\hat\alpha}}\Big(1 -
{4 \hat\alpha\over{l^2}}\Big)^{1\over 2}\Big]r^2.
\end{equation}
Hence we notice that the metric is real if,

\begin{equation}
\hat \alpha \le {l^2\over 4}.
\end{equation}

We shall restrict ourselves to $\hat \alpha$ which satisfy the above bound.
In this paper, we will primarily consider black holes in five dimensions $(n =
4)$. However, it is easy to extend the results of this section to higher dimensions.

The thermodynamic properties of the black hole will depend on
whether we consider the canonical ensemble (fixed charge $Q$) or
grand canonical ensemble (fixed potential $\Phi$). The equilibrium
temperature $T$ can identified from the period $\beta$ of the
Euclidean time of the metric (\ref{cgbsol}), which in five
dimensions is given by

\begin{equation}
\beta = {2\pi r_+( r_+^2 + 2\hat \alpha )\over{r_+^2 + 2 r_+^4/l^2 - q^2/r_+^2}}
\label{tempo}.
\end{equation}

As it will be useful for us to write thermodynamic quantities in terms of
dimensionless quantities, we define

\begin{equation}
\bar r = {r\over l}, \bar\alpha = {\hat\alpha\over l^2}, \bar q = {q\over l^2},
\bar m = {m\over l^2}.
\end{equation}
In terms of these quantities, (\ref{tempo}) can be expressed as

\begin{equation}
\beta = {2\pi l \bar r( \bar r^2 +
2\bar \alpha )\over{\bar r^2 + 2 \bar r^4 - \bar q^2/\bar r^2}}.
\label{temp}
\end{equation}

\bigskip

\subsection{Grand canonical ensemble}\label{gcan}

\bigskip

In the grand canonical ensemble, with fixed potential $\Phi$, the free energy can be
computed from the Euclidean continuation of the action (\ref{act}). We obtain
the action (subtracting the AdS background) as
\begin{equation}
I_{\rm gc} = - {\omega_3 l^2\beta\over{\kappa_5 (\bar r^2 +2 \bar \alpha)}}
\Big[ \bar r^6 + (18 \bar \alpha -1 + 4\Phi^2/3 )\bar r^4 +
3 \bar \alpha (1 - 8 \Phi^2/3)\bar r^2 - 6 \bar \alpha^2\Big],
\label{gcac}
\end{equation}
where $\beta$ is inverse of $T$ expressed in terms of potential
\begin{equation}
\beta = {2 \pi l(\bar r^2 + 2\bar\alpha)\over{\bar r(1 - 4 \Phi^2/3 + 2\bar r^2)}}.
\label{gcbeta}
\end{equation}

\begin{figure}[t]
\begin{center}
\begin{psfrags}
\psfrag{e}[][]{$\Phi^2$}
%\psfrag{q}[][]{$\bar{q}^2$}
\psfrag{f}[][]{$\bar{\alpha}$}
\psfrag{a}[][]{I}
\psfrag{b}[][]{II}
\psfrag{c}[][]{III}
\psfrag{d}[][]{IV}
%\psfrag{3/4}[][]{$3/4$}
%\psfrag{0.03}[][]{$0.03$}
\epsfig{file=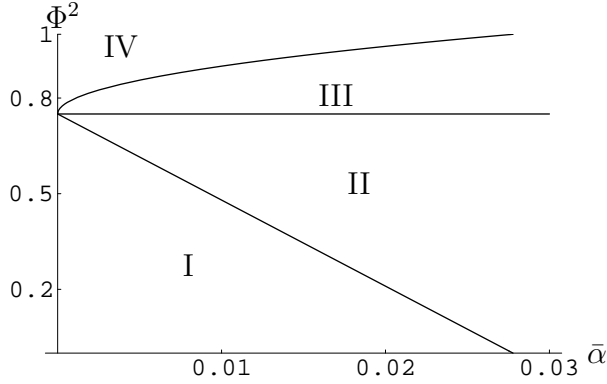, width= 8cm,angle=0}
\end{psfrags}
\vspace{ .1 in }
\caption{The curves in the (${\bar\alpha}$-$\Phi^2$) plane
separating various regions with different behaviours of black holes.}
\label{alphaqphia}
\end{center}
\end{figure}

It will be important for us to find out the number of turning points of
$\beta(\bar r)$ as we vary $\bar \alpha$ and $\Phi$. First of all, the
nature of
$\beta(\bar r)$ depends crucially on the value of $\Phi^2$. For $\Phi^2 >
3/4$,
$\beta(\bar r)$ blows up at $\bar r^2 = (4 \Phi^2/3 -1)/2$. Consequently,
the
temperature is zero. Following \cite{cejmo, cejmt}, we would like to identify this with an extremal hole.
For $\bar r$ less than this value, $\beta (\bar r)$ becomes negative. It
can
easily be checked that as long as $\Phi^2 > 3/4$, regardless of the
value of $\bar \alpha$, there is no turning point of $\beta(\bar r)$. If,
on the
other hand,
$\Phi^2 = 3/4$, $\beta$ diverges at $\bar r = 0$ and goes to zero for
large $\bar r$. Now, to have turning points,
$\partial \beta/\partial \bar r = 0$. This gives
\begin{equation}
\bar r_{1,2}^2 = {1\over{12}}\Big(3 - 36 \bar \alpha - 4 \Phi^2 \pm
{\sqrt{(-3 +12 \bar \alpha + 4 \Phi^2)(-3 + 108\bar\alpha + 4 \Phi^2)}}\Big).
\end{equation}
From here, it follows that in order to have real roots, $\bar \alpha$ should
{\it not} lie within the window
\begin{equation}
{1\over{36}}\Big(1 - {4 \Phi^2\over 3}\Big) \le \bar\alpha \le
{1\over{4}}\Big(1 - {4 \Phi^2\over 3}\Big).
\label{alphab}
\end{equation}
However, it is easy to check that for $\bar \alpha \ge
{1\over{4}}(1 - {4 \Phi^2\over 3})$, $\bar r_{1,2}^2$ are negative while
$\bar \alpha \le {1\over{36}}(1 - {4 \Phi^2\over 3})$, $\bar r_{1,2}^2$ are
positive. Hence, we have the following picture. For $\Phi < {{\sqrt3}/ 2}$,
$\beta$ has two turning points only if

\begin{equation}
\bar \alpha \le {1\over{36}}(1 - {4 \Phi^2\over 3}).
\label{alphac}
\end{equation}

\begin{figure}[t]
\begin{center}
\begin{psfrags}
\psfrag{p1}[][]{$\Phi^2=1.0$}
\psfrag{a1}[][]{$\bar{\alpha}=0.0$}
\psfrag{p2}[][]{$\Phi^2=0.6$}
\psfrag{a2}[][]{$\bar{\alpha}=0.0$}
\psfrag{p3}[][]{$\Phi^2=0.2$}
\psfrag{a3}[][]{$\bar{\alpha}=0.01$}
\psfrag{p4}[][]{$\Phi^2=0.8$}
\psfrag{a4}[][]{$\bar{\alpha}=0.02$}
\psfrag{a}[][]{(a)}
\psfrag{b}[][]{(b)}
\psfrag{c}[][]{(c)}
\psfrag{d}[][]{(d)}
\epsfig{file=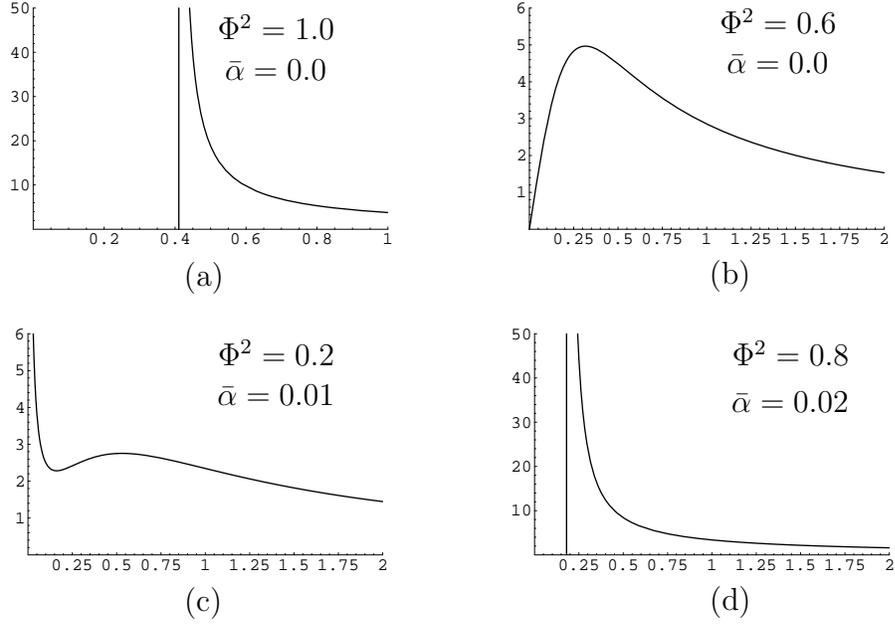, width= 12cm,angle=0}
\end{psfrags}
\vspace{ .1 in }
\caption{Plots of $\beta$ {\it vs} $\bar{r}$ for various values of
$(\Phi^2,\bar{\alpha})$.}
\label{phialpha}
\end{center}
\end{figure}

The above features of $\beta(\bar r)$ can be nicely summarized is a
($\Phi^2-\bar\alpha$) diagram. This is shown in Figure \ref{alphaqphia}.
The region satisfying (\ref{alphac}) is the region I in the figure. So,
here $\beta(\bar r)$ has two turning points. However, note that for
$\Phi^2 <
4/3$ and $\alpha =0$, $\beta(\bar r)$ has only one turning point at
non-zero
$\bar r$. In the rest of the regions namely II, III and IV,
there are no turning points of $\beta(\bar r)$. However, as in I, in
region II, $\beta(\bar r)$ diverges at $\bar r =0$ while in regions III
and IV,
$\beta(\bar r)$ blows up at finite non-zero values of $\bar r$. There are
other
differences in these four regions (particularly when the free energies of
the black holes are considered). This is what we discuss in the next
paragraph. Various representative plots of the $\beta$ versus $\bar{r}$ for
all these regions are shown in Figure \ref{phialpha}.

%We list here various solutions corresponding to the regions in
%Figure \ref{alphaqphia}.
%
%\begin{itemize}
%
%\item
%Above the straight line we
%have one real positive solution for $\bar{r}$ and below it there
%are a maximum of three real positive solutions at a particular
%temperature.
%
%\item
%For $\bar{\alpha}=0$ i.e. along the vertical axis we have only a
%maximum of two real positive solutions for $\Phi^2< 3/4$.
%
%\item
%Thermal AdS exists at all temperatures for all values of the parameters
%$\bar{\alpha}$ and $\Phi^2$.
%
%\end{itemize}
%
%Various representative plots of the $\beta$ versus $\bar{r}$ for the
%various regions are shown in Figure \ref{phialpha}.

\begin{figure}[t]
\begin{center}
\begin{psfrags}
\psfrag{a}[][]{(a)}
\psfrag{b}[][]{(b)}
\psfrag{c}[][]{(c)}
\psfrag{d}[][]{(d)}
\psfrag{tc}[][]{$T_c$}
\psfrag{t2}[][]{$T_2$}
\psfrag{t1}[][]{$T_1$}
\psfrag{i}[][]{$I$}
\psfrag{ii}[][]{$II$}
\psfrag{iii}[][]{$III$}
%\psfrag{c}[][]{(c)}
%\psfrag{d}[][]{(d)}
\epsfig{file=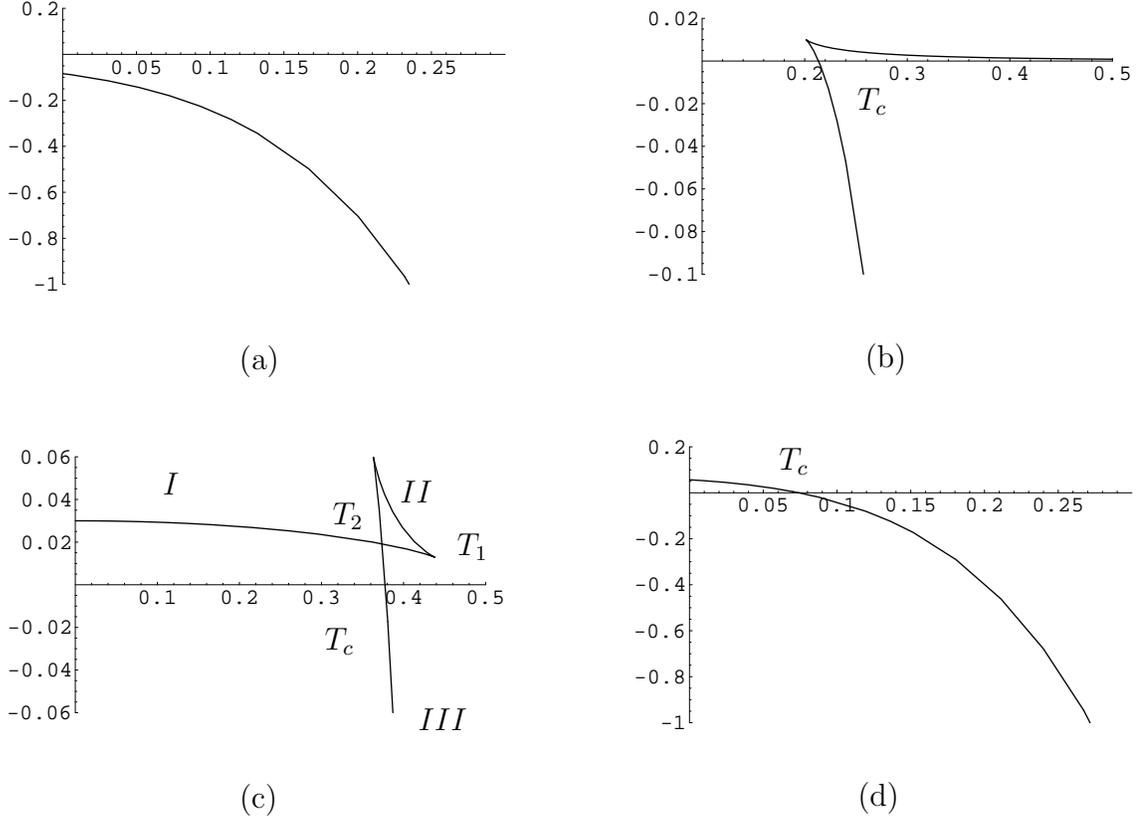, width= 15cm,angle=0}
\end{psfrags}
\vspace{ .1 in }
\caption{Free energy $W$ as a function of $T$. (a), (b), (c), (d) correspond to values of $\Phi^2$
and $\bar{\alpha}$ of Figure(\ref{phialpha}) }
\label{ftphi}
\end{center}
\end{figure}

\bigskip

\noindent
{\bf Free Energy:} The free energy can be obtained from (\ref{gcac}) as $W
= I_{\rm gc}/\beta$. For different values of the parameters, $W$ has
been plotted as a function of temperature in Figure \ref{ftphi}.
In this figure, (a) corresponds to the parameter values where we have only
one stable black hole solution.  This is also the situation in the case of (d).
However, there is
a distinct difference in their phase structures as is evident from the plots.
While the black hole phase has lower free energy than thermal AdS in
(a) for all temperatures, there is a Hawking-Page transition at $T_c$ in
(d). This difference can easily be located in ($\Phi^2-\bar\alpha$) diagram
in Figure \ref{alphaqphia}. In the region IV of this figure, black hole at
$T=0$ has less energy than the thermal AdS and hence is stable. However, in
regions II and III, a hole at $T=0$ is in a metastable phase while AdS is
the stable one. Hence this hole would decay to AdS by radiating away its
energy. The line separating III and IV represents hole with zero free
energy at $T=0$. The equation of this line as a function of $\bar{\alpha}$ and 
$\Phi^2$ is obtained by setting $W(\bar{r}, \bar{\alpha},\Phi^2)=0$, where 
$\bar{r^2}=1/2(1-4\Phi^2/3)$. Note that this also means that on this line the Hawking-Page temperature
vanishes.

Now returning back to Figure \ref{ftphi}(b), we see that
at low temperatures there is no black hole phase.
Two black hole phases appear as we increase the
temperature. The small one turns out to be  unstable and the large one
undergoes a Hawking Page transition at $T_c$. Note that Figure
\ref{ftphi}(c) is similar to the one we found in our previous paper
\cite{dmms} (i.e for $\Phi^2=0,\bar{\alpha}\ne 0$).
As in \cite{dmms}, we have therefore the following scenario. At low
temperature, free energy has only one branch (branch I). However, as
temperature is increased, two new branches (branch II and III) appear.
Branch II meets branch
I at a certain temperature ($T_1$) and they both disappear. On the other hand,
branch III continues to decrease, cuts branch I at a particular temperature
($T_2$) and becomes negative at temperature $T_c$. These three
branches represent small, intermediate and large black hole. Out of these
three, the intermediate black hole is unstable with negative specific heat,
while the rest are classically
stable. As in \cite{dmms}, we get two first order phase transitions (HP1, HP2).
HP1 is a transition between small and large black hole at temperature $T_2$
and the other (HP2) is the
usual transition between AdS to large black hole as $T_c$.

\begin{figure}[t]
\begin{center}
\begin{psfrags}
\psfrag{a}[][]{(a)}
\psfrag{b}[][]{(b)}
\psfrag{c}[][]{(c)}
\psfrag{d}[][]{(d)}
\psfrag{tc}[][]{$T_c$}
\psfrag{t2}[][]{$T_2$}
%\psfrag{t1}[][]{$T_1$}
%\psfrag{i}[][]{$I$}
%\psfrag{ii}[][]{$II$}
%\psfrag{iii}[][]{$III$}
%\psfrag{c}[][]{(c)}
%\psfrag{d}[][]{(d)}
\epsfig{file=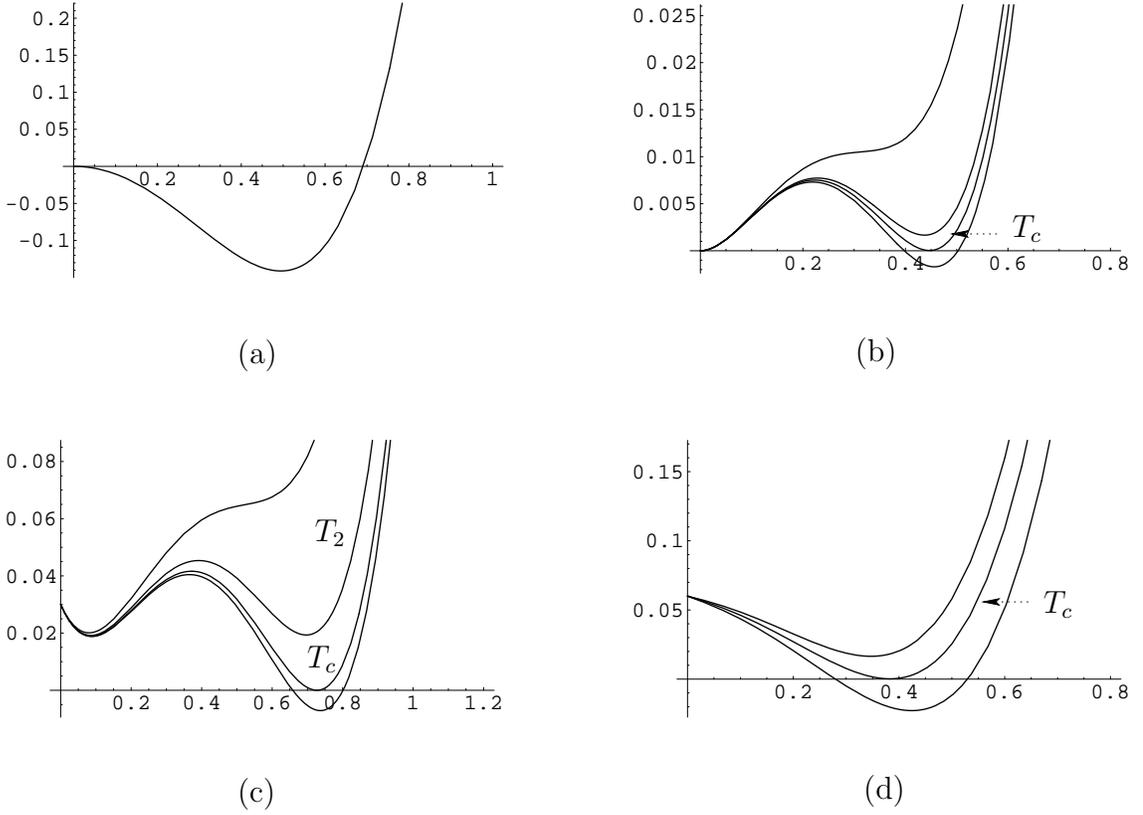, width= 15cm,angle=0}
\end{psfrags}
\vspace{ .1 in }
\caption{Landau function ${\cal W}$ {\it vs.} $\bar r$ for
different temperatures. (a), (b), (c), (d) correspond to values of $\Phi^2$
and $\bar{\alpha}$ of Figure(\ref{phialpha})}
\label{lphi}
\end{center}
\end{figure}

It is easy to construct a Landau
function which represents the behaviour of the free energy around the critical points. Identifying
$\bar r$ as the order parameter, this function can be written as
\begin{equation}
{\cal {W}} (\bar r, T) = {\omega_3 l^2\over {\kappa_5}}\Big[ 3 {\bar r}^4 - 4 \pi l T {\bar r}^3
+ ( 3 - 4 \Phi^2) {\bar r}^2 - 24 \pi \bar \alpha l T \bar r + 3 \bar \alpha\Big].
\end{equation}
Notice that this expression reduces to the one in \cite{BM} when we set $\bar \alpha$ to zero and
to the one in \cite{dmms} as we set $\Phi$ to zero.
It can be checked that at the saddle point of this function, we get the temperature (given by
the inverse of the expression in (\ref{gcbeta})). Substituting back the temperature in
${\cal{W}}$, we get the free energy $W$. We have plotted the Landau function in
Figure \ref{lphi}. Consider $(c)$ in Figure \ref{lphi} in particular. This is for
$\Phi^2 = 0.2, \bar\alpha = 0.01$. Clearly, for $T < T_2$, the global minimum occurs for
small $\bar r$, representing a stable small black hole phase. At $T = T_2$, small and big
black hole co-exist. At even higher temperature, big black hole represents the
stable phase. However, all these phases are meta-stable below $T < T_c$ if we include
thermal AdS (representing the horizontal line with ${\cal{W}} = 0$). Big holes then are only
stable beyond $T_c$.
Note, that for $(a)$ in Figure 4, black hole is always the stable phase while for
$(b)$, there is a crossover from thermal AdS to black hole at $T_c$. Finally,
Figure $(d)$ is similar to $(a)$ except that the $\bar r = 0$ hole has
more energy than the thermal AdS.
To this end, we note that
since $W = E - T S - \Phi Q$, we get the energy, entropy and charge as \begin{eqnarray}
&&E = \Big({\partial I_{\rm gc}\over{\partial\beta}}\Big)_\Phi - {\Phi\over
\beta}\Big(
{\partial I_{\rm gc}\over{\partial\Phi}}\Big)_\beta = M \nonumber\\
&& S = \beta \Big({\partial I_{\rm gc}\over{\partial\beta}}\Big)_\Phi - I_{\rm gc}=
{4 \pi l^3 \omega_3 \bar r (\bar r^2 + 6 \bar \alpha)\over{\kappa_5}}
\nonumber\\
&& Q = - {1\over \beta}\Big({\partial I_{\rm gc}\over{\partial\Phi}}\Big)_\beta,
\label{gcthermo}
\end{eqnarray}
where expressions for $Q, M$ in terms of $q, m$ respectively were defined earlier. It can be
checked that the first law of thermodynamics
$dE = T dS + \Phi dQ$ is satisfied.

\bigskip

\subsection{Canonical ensemble}\label{can}
\begin{figure}[t]
\begin{center}
\begin{psfrags}
\psfrag{a}[][]{$\bar{q}^2$}
\psfrag{b}[][]{$\bar{\alpha}$}
\psfrag{c}[][]{$I$}
\psfrag{d}[][]{$II$}
\epsfig{file=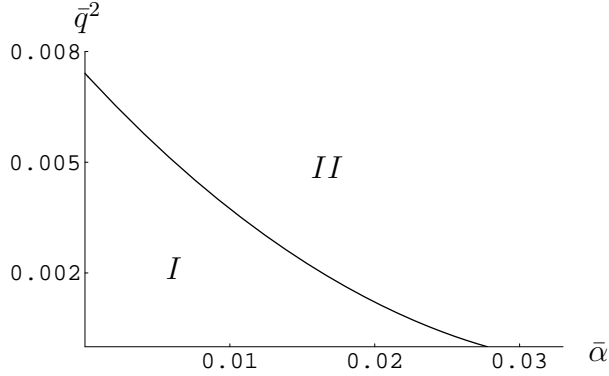, width= 8cm,angle=0}
\end{psfrags}
\vspace{ .1 in }
\caption{The curve in
$\bar{\alpha}-\bar q^2$ plane separating regions with various number of solutions.}
\label{alphaqphib}
\end{center}
\end{figure}

\bigskip

We now consider the system in canonical ensemble where the charge $q$ is kept fixed.
We first note from the expression of the temperature (\ref{tempo}) that $T$ is
non-negative if
\begin{equation}
\bar r^2 + 2 \bar r^4- {\bar q^2\over {\bar r^2}} \ge 0.
\end{equation}
When the equality is saturated, the temperature is zero and we call this an
extremal black hole. Denoting mass and the scaled horizon radius as $\bar
m_e$ and $\bar r_e$ respectively, we see that the following relation is
satisfied:
\begin{equation}
\bar m_e = \bar \alpha + {\bar r_e^2\over 2} + {3 \bar q^2\over{2\bar r_e^2}}.
\end{equation}

\begin{figure}[t]
\begin{center}
\begin{psfrags}
\psfrag{q1}[][]{$\bar{q}^2=0.01$}
\psfrag{a1}[][]{$\bar{\alpha}=0.0$}
\psfrag{q2}[][]{$\bar{q}^2=0.002$}
\psfrag{a2}[][]{$\bar{\alpha}=0.0$}
\psfrag{q3}[][]{$\bar{q}^2=0.002$}
\psfrag{a3}[][]{$\bar{\alpha}=0.005$}
\psfrag{q4}[][]{$\bar{q}^2=0.008$}
\psfrag{a4}[][]{$\bar{\alpha}=0.015$}
\psfrag{a}[][]{(a)}
\psfrag{b}[][]{(b)}
\psfrag{c}[][]{(c)}
\psfrag{d}[][]{(d)}
\epsfig{file=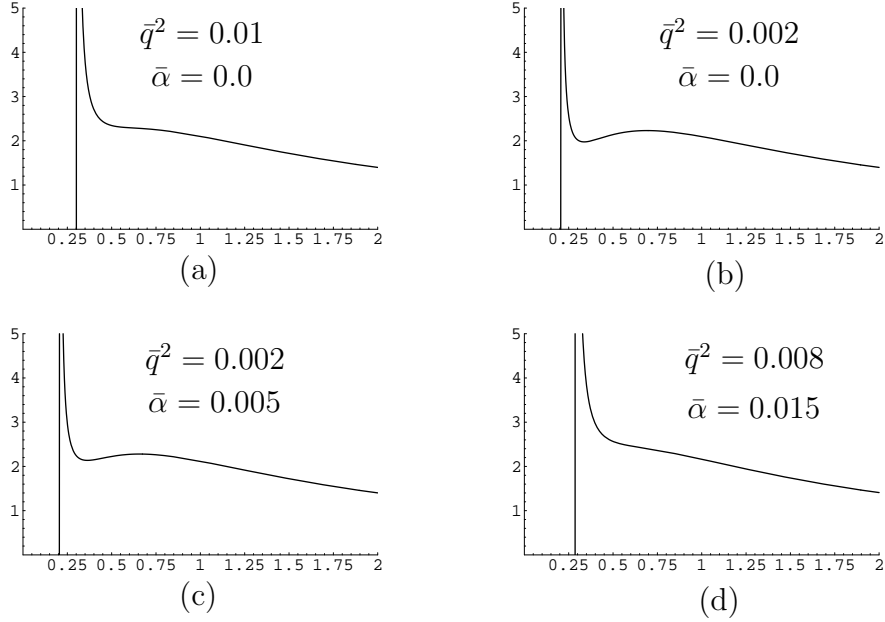, width= 12cm,angle=0}
\end{psfrags}
\vspace{ .1 in }
\caption{Plots of $\beta$ {\it vs} $\bar{r}$ for various values of
$(\bar{q}^2,\bar{\alpha})$.}
\label{qalpha}
\end{center}
\end{figure}

Like in the previous case of fixed potential, we now identify the relevant
regions in the ($\bar{\alpha}$-$\bar{q}^2$) plane.
The curve separating the regions
for various number of positive solutions for $\bar{r}$ is given by the
following parametric equations in $\bar{r}$:
\begin{eqnarray}{\label{qbaralphab}}
\bar{q}^2&=&\frac{1}{15}\left(6\bar{r}^6-\bar{r}^4\right),\\ \nonumber
\bar{\alpha}&=& \frac{5}{3}\left(\frac{\bar{r}^2-3\bar{r}^4}{18\bar{r}^2+2}
\right).
\end{eqnarray}
The curve is shown in Figure \ref{alphaqphib}.
It can easily be checked that for any point in region II,
there is one real positive root for $\bar{r}$ at any
temperature. Below this, that is in region I, there is a maximum of three.
Furthermore, unlike the fixed potential case,
the vertical $\bar{q}^2$-axis i.e. for $\bar{\alpha}=0$ we also have a
maximum of three real positive solutions as long as $\bar{q}^2 <
\bar{q}^2_c(=1/135)$.
We also notice that thermal AdS exists only when $\bar{q}^2=0$. The corresponding
$\beta$-$\bar{r}$ plots for these regions are shown in Figure \ref{qalpha}.

\bigskip

\noindent
{\bf Free Energy:} We now compute the action (\ref{act}) in the fixed
charge ensemble. After
properly adding a boundary charge and subtracting the contribution to the
extremal background, we get
\begin{equation}
I_{\rm c} = {\omega_3 l^2 \beta\over{\kappa_5}}\Big[ \bar r^2 - \bar r^4 +
{5 \bar q^2\over {\bar r^2}} + {8 {\bar \alpha} ({\bar q}^2 - {\bar r}^4 - 2 {\bar r}^6)
\over{ {\bar r}^2 ({\bar r}^2 + 2 \bar \alpha )}}
- {3\over 2} \bar r_e^2 - {9 \bar q^2\over{2 \bar
r_e^2}}\Big],
\end{equation}
where $\beta$ is
\begin{equation}
\beta = { 2 \pi l (\bar r^2 + 2\bar\alpha)\bar r\over{\bar r^2 + 2 \bar r^4 -
\bar q^2/\bar r^2}}.
\end{equation}

\begin{figure}[t]
\begin{center}
\begin{psfrags}
\psfrag{(a)}[][]{$(a)$}
\psfrag{(b)}[][]{$(b)$}
\psfrag{(c)}[][]{$(c)$}
\psfrag{(d)}[][]{$(d)$}
\psfrag{f}[][]{$\bar T $}
%\psfrag{ii}[][]{(B)}
\epsfig{file=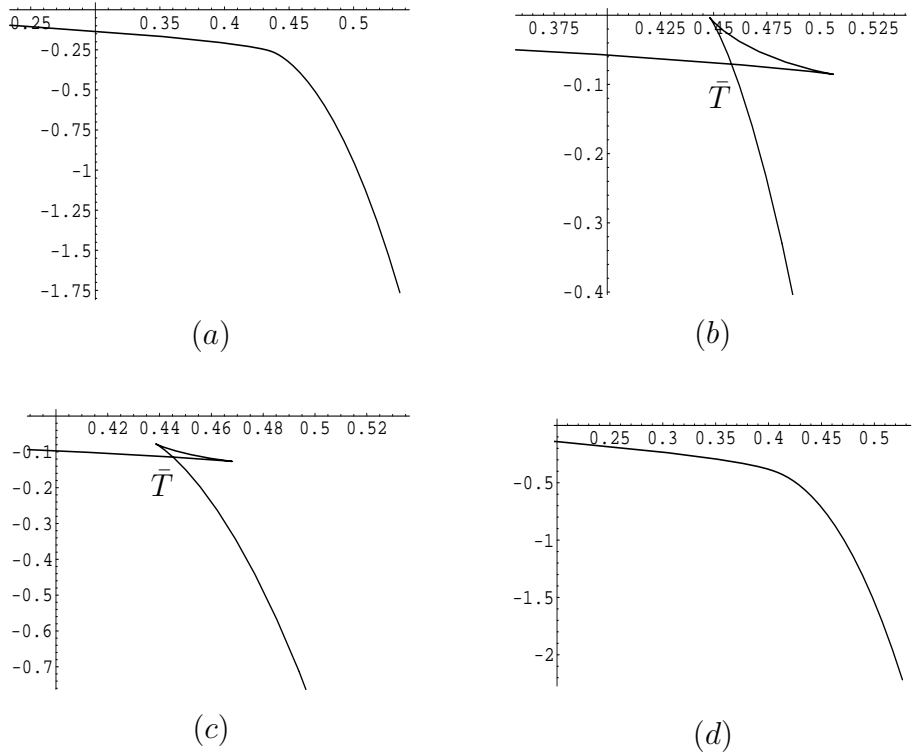, width= 12cm,angle=0}
\end{psfrags}
\vspace{ .1 in }
\caption{The free energy in canonical ensemble as a function of temperature. (a), (b), (c), (d) represent
plots for values of ${\bar q}^2$ and $\bar \alpha$ of Figure (5).}
\label{onetwothreefour}
\end{center}
\end{figure}

The free energy can therefore be obtained as $F = I_{\rm c}/\beta$.
Behaviours of free energy for different values of $(q^2, \bar\alpha)$
are shown in Figure \ref{onetwothreefour}.
We first of all note that, in fixed charge ensemble, black holes with negative free energies are
always the stable
compared to thermal AdS. Secondly, in $(a)$ and $(d)$, we see that given any temperature, there is
a
single black hole phase, while in $(b)$ and $(c)$, there can atmost be three phases.
We call them
small, big and intermediate black hole phases.
We find that at a certain temperature, which we call $\bar T$ later, there
is a first order phase transition from small to big black hole phase. On the other hand, the
intermediate black hole is an unstable phase with negative specific heat. It can be shown by
comparing $(b)$ and $(c)$, that $\bar T$ decreases as $\bar \alpha$ increases.

\begin{figure}[t]
\begin{center}
\begin{psfrags}
\psfrag{(a)}[][]{$(a)$}
\psfrag{(b)}[][]{$(b)$}
\psfrag{(c)}[][]{$(c)$}
\psfrag{(d)}[][]{$(d)$}
\psfrag{aa}[][]{$\bar T$}
\psfrag{ab}[][]{$\bar T$}
\epsfig{file=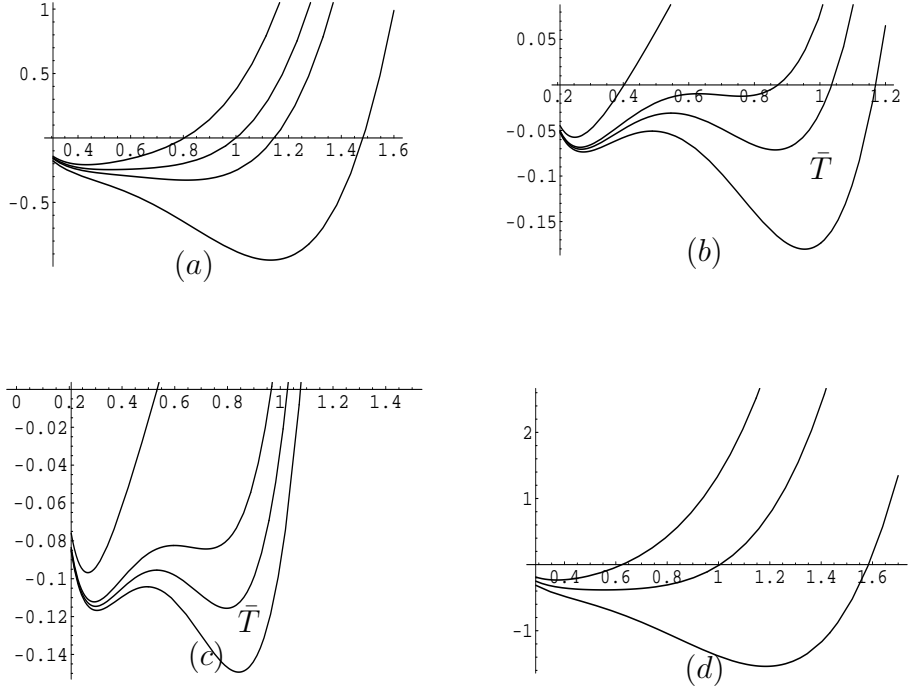, width= 12cm,angle=0}
\end{psfrags}
\vspace{ .1 in }
\caption{The Landau Function ${\cal{F}}$ as a function of $\bar r$ for different temperatures. The
critical
temperature at which there is a transition between the small black hole to the large black
hole is shown by
$\bar T$ in the figures. Finally, (a), (b), (c), (d) represent plots for
values of ${\bar q}^2$ and $\bar
\alpha$ of Figure \ref{qalpha}.}
\label{oneonetwotwo}
\end{center}
\end{figure}

Finally, the Landau function can be constructed as before. It is given by
\begin{equation}
{{\cal F}}(\bar r, T) = {\omega_3l^2\over \kappa_5}\Big[3 {\bar r}^4 - 4 \pi l T {\bar r}^3  +
3 {\bar r}^2 - 24 \pi l {\bar \alpha} T \bar r + {3 {\bar q}^2\over {{\bar r}^2}} - {9 {\bar q}^2\over{2
{{\bar r}_e}^2}} - {3 {{\bar r}_e}^2\over 2}\Big].
\end{equation}
It can be checked that, at the saddle point, it reproduces correct temperature $T$
and  the
free energy
$F$. A plot of this function for different temperatures is shown in Figure
\ref{oneonetwotwo}. As can be seen in $(a)$, for high $\bar q$, there is
a single black hole phase for all temperatures. As we reduce $\bar q$ beyond certain value, two new
black hole phases appear in $(b)$ for a certain range of temperature.
Above and below this range there is only one black hole solution.
When the temperature is within this range, for $T < \bar T$, the small
black hole is favoured. Otherwise,
big black hole is the stable one. There are degenerate minima at $T = \bar T$ representing phase co-existence.

Now as we turn on $\bar \alpha$, we get $(c)$ for low values of
$\bar q, \bar \alpha$. This is similar to $(b)$ except that the critical
temperature $\bar T$ reduces with $\bar \alpha$. Again, for
large $\bar \alpha$, we get $(d)$ which is qualitatively similar to that
of $(a)$.

Finally, since $F = E - TS$, we have
\begin{eqnarray}
&&E = \Big({\partial I_{\rm c}\over{\partial\beta}}\Big)_Q = M - M_e\nonumber\\
&&S = \beta \Big({\partial I_{\rm c}\over{\partial \beta}}\Big)_Q - I_{\rm c} =
{4 \pi l^3 \omega_3 \bar r (\bar r^2 + 6 \bar \alpha)\over{\kappa_5}}\nonumber\\
&&\Phi = {1\over\beta} \Big({\partial I_{\rm c}\over{\partial Q}}\Big)_\beta
= {\sqrt{3\over 4}} \Big({\bar q\over {\bar r^2}} - {\bar q\over {\bar r_e^2}}\Big).
\end{eqnarray}
It can be checked that they satisfy the first law of thermodynamics
$dE = T dS + (\Phi -\Phi_e)dQ$.

%%%%%%%%%%%%%%%%%%%%%%%%%%%%%%%%%%%%%%%%%%%%%%%%%%%%%%%%%%%%%%%%%%

\section{Dual Matrix Model}\label{boundary}

Classical Type IIB supergravity is dual to ${\cal N}=4$ $SU(N)$
gauge theory at strong coupling. The phases of the gauge theory in
this regime of the coupling is very hard to analyze. Assuming the
AdS/CFT correspondence to be true one can on the other hand get some
understanding from the supergravity computations, that are more
tractable. Following these lines, Witten has argued in
\cite{wittenone} that the Hawking-Page transition in gravity
corresponds to the confinement/deconfinement transition on the
boundary. This is the philosophy that we will follow.

In this section, using the AdS/CFT correspondence we write down the
dual matrix model corresponding to the $R$-charged black holes that
were studied in the earlier sections. The matrix model is
constructed phenomenologically so that it captures the phases of the
black holes. The results from the weak coupling computations
available in the literature will be used as a guideline in this
prescription as was also followed in \cite{dmms}. We begin with a
brief review of some of the results that we will need for the finite
temperature gauge theory at weak coupling.

\subsection{Weak coupling}

${\cal N}=4$, $SU(N)$ gauge theory at zero and weak couplings has
been analyzed by various authors (see for example \cite{bs,ammpr}).
For large $N$, when the t'Hooft coupling $\lambda=g^2_{YM}N$ is
small or zero, some of the results are explicitly known.
Specifically, when $\lambda=0$, it was shown that the boundary
gauge theory at finite temperature on $S^3\times S^1$ undergoes a
phase transition that can be identified as the``deconfinement"
transition. $S^3\times S^1$ is a compact manifold and thus allows
only colour singlet states by the Gauss law constraint. Though non
singlet states are never possible, there are various indications
that this transition mimics the deconfinement transition in gauge
theories. One of the indications is that there is a jump of the free
energy from order $N^0$ to order $N^2$ while another is a discrete
change in expectation value of the Polyakov loop.

Dimensional reduction of the ${\cal N}=4$ theory on $S^3\times S^1$
leaves only one massless mode, namely, the zero mode of $A_0$. One
can thus write down an effective action by integrating out all
massive modes. The resulting model with the gauge fixing conditions,
$\partial_i A_i=0$ and $\partial_t \alpha=0$, is a zero dimensional matrix
model given by,

\beqal{eff} Z&=&\int DU e^{S_{eff}(U)}, \non
U&=&e^{i\beta\alpha}\mbox{\hspace{.1in},\hspace{.1in}}
\alpha=1/\omega_3 \int_{S^3} A_0 , \eeqa

\noindent where $\omega_3$ denotes the volume of $S^3$. Apart from the free
energy, whose $N$ dependence reflects the phase of the gauge theory,
one can also define a Polyakov loop that acts as an order parameter
for the deconfinement transition. The Polyakov loop defined along
the time circle of $S^3\times S^1$ is 
$(1/N)\Tr P \exp(i\int_0^{\beta} dt A_0)$,
which is nothing but $(1/N) \Tr(U)$ as can be seen from (\ref{eff}).
It can be shown that the expectation value of the Polyakov loop does
indeed vanish at low temperatures and picks up a non-zero value
above some finite temperature, $T_H$ which may be called the
Hagedorn or the "deconfinement" transition temperature. 

The explicit form of the partition function is given by,

\beqal{singlep}
Z&=&\int dU \exp \left[\sum_{n=1}^{\infty}
\frac{1}{n}z(x^n)\left(\Tr(U^n)\Tr(U^{-n})\right)
\right]\\\nonumber
z(x^n)&=&z_V(x^n)+z_S(x^n)+ (-1)^{n+1}z_F(x^n)
\mbox{\hspace{.1in};\hspace{.1in}} x=e^{-\beta}.
\eeqa

\noindent Introducing the density of eigenvalues for $U$ and defining $\rho=
(1/N)\Tr(U)$ one can write down the effective action as
$S_{eff}$ (see \cite{ammpr} for details). In the large $N$
approximation, the various phases are given by the solutions of the
saddle point equations of motion. In terms of the density of
eigenvalues, the above transition is reflected by a jump of $\rho$
from zero to a nonzero value below and above $T_H$ respectively,
where $T_H$ is given by $z(x)=1$.

These features are some what modified when a small value of the coupling
$\lambda$ is turned on. It is possible to write $S_{eff}$ only in terms of
powers of $\Tr(U)$ by using the
saddle point equations. An effective action containing  only the quartic
interactions may be written as \footnote{This model is obtained by keeping
terms upto ${\cal O}(\lambda^2)$ in the effective action. In the large $N$
limit such terms come from three loop computations. It also determines the
sign of $b$. In \cite{ammpr} the phases for both the signs of $b$ have been studied.
An explicit computation for pure Yang Mills theory on a three sphere at finite
temperature shows that $b$ is positive, implying that the transition
is of first order at weak coupling \cite{Aharony:2005bq}.},

\beqal{weakeff}
Z(\lambda,T)=\int dU \exp\left[a(\lambda,T)\Tr(U) \Tr(U)^{\dagger}+\frac{b}{N^2}
(\lambda,T)\left(\Tr (U) \Tr (U)^{\dagger}\right)^2\right].
\eeqa

The equations of motion resulting from (\ref{weakeff}) are,

\begin{eqnarray}\label{saddle}
a\rho+2b\rho^3&=&\rho \mbox{\hspace{0.7in}} 0\le \rho \le \frac{1}{2}
\nonumber\\
&=&\frac{1}{4(1-\rho)} \mbox{\hspace{0.2in}} \frac{1}{2}\le \rho\le 1.
\end{eqnarray}

\noindent The matrix model (\ref{weakeff}) undergoes two different phase
transitions as a function of temperature. One is a first order
transition like the zero coupling case, when $b>0$. The other is a
third order transition for which the eigenvalue distribution
goes from the gapless phase for $0\le \rho \le \frac{1}{2}$,
to a phase with a gap for $\frac{1}{2}\le \rho\le 1$.

\subsection{Strong coupling and comparison to gravity}

If one assumes validity of (\ref{weakeff}) in the strong coupling
regime (where $a$ and $b$ are some complicated functions of
$\lambda$) one may hope to map these phases to those of gravity
obtained in the supergravity approximation. It was shown that this
simplified model indeed possesses the thermodynamic behaviour
expected from gravity when $a<1$ and $b>0$ \cite{aglw}. A new
feature that is not visible in the supergravity approximation is the
transition corresponding to the third order transition that was
mentioned in the earlier paragraph. It was conjectured that this
should correspond to black hole string transition
\cite{Horowitz:1996nw}. It should however be noted that the matching is
only qualitative and the validity of the effective action
(\ref{weakeff}) in the strong coupling regime is only limited to the
regions around the critical points.

The phase structure gets modified once we include ($\alpha^{'}$)
corrections. In addition to the large stable black hole, there is
also a small stable black hole solution along with an intermediate
unstable one. There are two phase transitions: (i) From the small to
the large black hole (HP1) at $T_2$ (ii) Between thermal AdS and the
large black hole (HP2) at $T_c$.

In \cite{dmms} we studied the matrix model that captures these
phases in gravity. Since the size of the small black hole is less
than $\sqrt{\alpha^{'}}$ while we include corrections only up to
first order of $\alpha^\prime$ we discussed both the following
possibilities.
If we do not trust this solution, the situation is same as that of
the case without higher derivative corrections and the $(a,b)$ model
(\ref{weakeff}) itself serves as the dual of gravity. With the
$\alpha^{'}$ corrections in the bulk it is also possible to see the
variations of $a$ and $b$ as a function of $1/\sqrt{\lambda}$. An
interesting point is to note that $b$ decreases as we decrease
$\lambda$.

On the other hand, if we trust this solution the minimal action for
the matrix model gets modified so that it can reproduce all the
features of gravity near the critical points, including the small
black hole. This modified action is given by,
\begin{equation}
S(\rho^2)= 2 N^2\left[ A_4 \rho^8  -  A_3 \rho^6 +  A_2 \rho^4 +
\left(\frac{1-2A_1}{2}\right) \rho^2 \right]. \label{action}
\end{equation}
The $A_i$'s are functions of $\lambda$ and $T$. The $(a,b)$ model is
recovered when $A_3$ and $A_4$ vanishes. The addition of
$\alpha^{'}$ corrections in gravity requires higher powers of $\rho$
to be added. In order that the model (\ref{action}) reproduces all
the features of gravity one also needs to restrict the values of
$A_i$'s within certain regions. There is however one saddle point
solution that does not have a counterpart in supergravity. This
solution is locally unstable and can possibly serve as some
stringy decay mode for the small black hole. In the following parts
of this section we will use (\ref{action}) so that it incorporates
the phases of charged black holes as discussed in section
(\ref{bulk}).

\subsection{Matrix model with chemical potential}

Once we turn on a non-zero chemical potential the coefficients of
the matrix models described above will depend on the chemical
potential. At zero coupling, {\it i.e} at $\lambda=0$, this
dependence is easy to determine. The partition function in this case
is given by (\ref{singlep}) with $z(x^n)$ modified as,

\beqal{chpot}
z(x^n)&=&z_V(x^n)+z_S(x^n)\cosh(\mu)+ (-1)^{n+1}z_F(x^n)\cosh(\mu/2).
\eeqa

\noindent This matrix model was studied in \cite{yy,TM} and similar to the case of $\mu=0$,
there is first order deconfinement transition at temperature $T_H$. Where
$T_H$ is given by $z(x)=1$.
There is a discontinuous change in $\rho$ from zero to a nonzero value.
Above $T_H$ the deconfined phase is preferred and has free energy
${\cal O}(N^2)$. Below $T_H$ the preferred confined phase has free energy
${\cal O}(1)$.

When a small $\lambda$ is turned on, to the quadratic order in
$\lambda$ one gets a term of the form
$\left(\Tr(U)\Tr(U^{-1})\right)^2$. One may thus be inclined to
propose a matrix model corresponding to the black holes with
chemical potential as was done earlier with the $(a,b)$ model for
the black holes zero chemical potential. However, the dependence of
the coefficients, $(a,b)$ on the chemical potential, though
obvious in the $\lambda=0$ case, is not easy to determine when
$\lambda \ne 0$. As was mentioned, the lowest correction is ${\cal
O}(\lambda^2)$ that involves a three loop computation\footnote{To
our knowledge this computation is not available in the literature.
We hope to return to this problem in future.}. On the other hand, an
alternative approach is to write down a model in the strong coupling
regime by comparing with gravity. Though we do not hope to determine
the exact dependence of the parameters on the chemical potential,
the restrictions on them so that the matrix model reproduces the
qualitative features of gravity can be inferred.

Let us begin with the case without $\alpha^{'}$ corrections. We will
consider the equation (\ref{saddle}), where now apart from $\lambda$
and $T$, $a$ and $b$ are also functions of the chemical potential
$\mu$.

When comparing with gravity, $a(\lambda,T,\mu)$ and
$b(\lambda,T,\mu)$ will be assumed to be positive for all values of $\mu$.
We have seen in section (\ref{gcan}) that there exists a critical
value of the potential ($\mu_c$) above
which there is only one solution. Below this there are a maximum of two
solutions.  The phase structure below the
critical potential is same as that of the uncharged case. This is
reproduced by the $(a,b)$ model when $a<1$ and $b>0$ as described in
\cite{aglw}.

\begin{figure}[t]
\begin{center}
\begin{psfrags}
\psfrag{a1}[][]{(A) $a(\mu,T) >1$}
\psfrag{a2}[][]{(B) $a(\mu,T) <1$}
\psfrag{T}[][]{$T=0$}
\psfrag{T1}[][]{$T> 0$}
%\psfrag{ii}[][]{(B)}
\epsfig{file=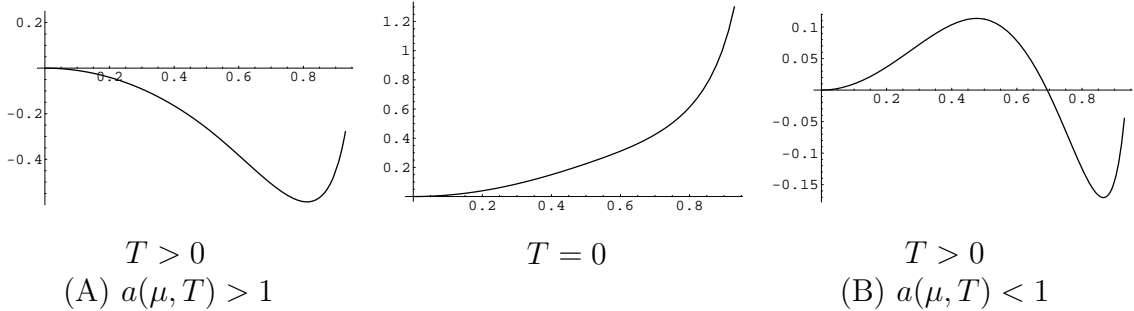, width= 15cm,angle=0}
\end{psfrags}
\vspace{ .1 in }
\caption{Plots of the matrix model potential {\it vs.} $\rho$ above $\mu_c$
showing the two possible situations at finite temperature.
At $T=0$ the extremal black hole decays into AdS, shown as a saddle point at
$\rho=0$.}
\label{fixedpotmat}
\end{center}
\end{figure}

From equation (\ref{gcbeta}), we see that the radius of the small black hole at
and above the critical potential, becomes zero and negative
respectively. The black hole with positive radius approaches an
extremal limit at $T=0$. We know that thermal AdS exists for all
temperatures and at $T=0$ the gauge theory is in the confined phase.
We thus have two solutions in gravity corresponding to the
confined phase. However the extremal black hole being
nonsupersymmetric will ultimately decay into AdS at zero
temperature\cite{cejmo, cejmt}. We do not expect to capture the
dynamics of this decay with the static potential given by the
$(a,b)$ model. In fact the zero temperature configuration with only
the AdS is not continuously connected to the finite temperature
$(a,b)$ model as we will see below.

At finite temperature, we always need to introduce an unstable
saddle point (a maximum). This follows from the observation that the
smooth potential given by the  $(a,b)$ model which gives two minima
(corresponding to thermal AdS and the large black hole in gravity)
always includes a maximum in between. For $\mu > \mu_c$ this maximum
corresponding to the unstable black hole ceases to show up in
gravity above. In the matrix model, we interpret this phenomenon as
follows. As $\mu$ increases beyond $\mu_c$ this unstable saddle
point in the matrix model enters the $\rho < 1/2$ region from $\rho
> 1/2$ region. Thus the region
beyond $\mu_c$ corresponds to the values of $a$ and $b$ when the
unstable small black hole of the $(a,b)$ model has already undergone
the third order Gross-Witten transition \cite{Gross:1980he,Wadia:1980cp}.
This imposes a constraint on the values of $a$ and $b$. Another
constraint comes from the fact that the energy of the large black
hole in this region is lower than thermal AdS for all temperatures.
At any finite temperature, the theory is thus always in the
deconfined phase. Both of these constraints are satisfied if we set
$b> 2(1-a)$. However, in light of these remarks, it is not possible
to tell whether the unstable saddle point has energy less or more
than that of AdS or whether it is at $\rho=0$ or away from it.  Thus
it gives rise to two possible scenarios depending on whether the
unstable saddle point is at $\rho=0$ or in the region $0 < \rho <
1/2$ as shown in Figure \ref{fixedpotmat}. In the following we
discuss these two scenarios separately .
\\

\noindent{\bf (A) Unstable maximum is at $\rho=0$ ($a(\mu \ge
\mu_c,T)\ge 1$):} In this case we define the critical potential
$\mu_c$ by $a(\mu_c,T) = 1$. Here the unstable saddle point at
$\rho=0$ does not correspond to thermal AdS but to the unstable
configuration not visible in gravity as mentioned above. Thermal AdS
does not feature in the plot. It has energy higher than the black
hole.  The condition  $b> 2(1-a)$ is automatically satisfied as $b$
is assumed to be positive.
\\

\noindent{\bf (B) Unstable maximum is at $0 < \rho < 1/2$ ($a(\mu
\ge \mu_c,T)<1$):} In this case the saddle point at $\rho=0$ is
thermal AdS. The unstable saddle point has energy higher than AdS.
Since the black hole has energy less than AdS, $a$ and $b$
correspond to values above the Hawking-Page transition. Also the
saddle point for the maximum is at $\rho < 1/2$. The critical
potential in this case is given by, $b(\lambda,T,\mu_c) =
2\left[1-a(\lambda,T,\mu_c)\right]$ or by the curve in the parameter
space of $a$ and $b$ that gives the Hawking-Page transition,
depending on whichever satisfies both the above conditions.
\\

\begin{figure}[h]
\begin{center}
\begin{psfrags}
\psfrag{i}[][]{I}
\psfrag{ii}[][]{II}
\psfrag{iii}[][]{III}
\psfrag{iv}[][]{IV}
\psfrag{1}[][]{(A)}
\psfrag{2}[][]{(B)}
\psfrag{a}[][]{$a=1$}
\psfrag{T=0}[][]{$T=0$}
\psfrag{T>0}[][]{$T> 0$}
%\psfrag{ii}[][]{(B)}
\epsfig{file=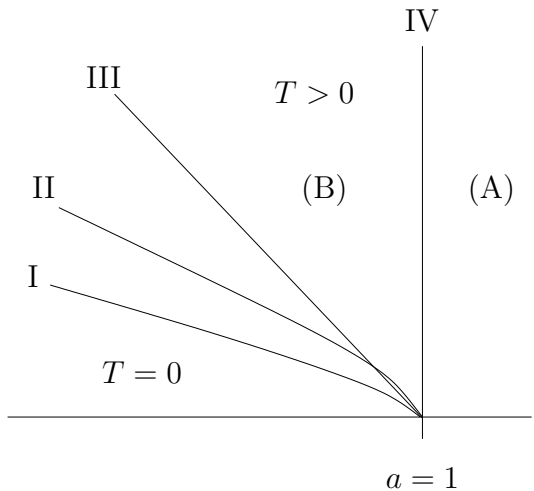, width= 7cm,angle=0}
\end{psfrags}
\vspace{ .1 in }
\caption{The $a-b$ parameter space for $\mu> \mu_c$.
The zero temperature region is below I and the finite temperature region
is beyond III or II. Below I there
is only one saddle point at $\rho=0$ corresponding to thermal AdS.
On II the Hawking Page transition occurs. III is the Gross-Witten transition
line. IV is the line, $a=1$}
\label{ab}
\end{center}
\end{figure}

At this point we recall the plot of \cite{aglw} showing the various regions
in the ($a-b$) parameter space. This is shown in Figure \ref{ab}.
For $\mu>\mu_c$ we have the following situation: At $T=0$ the only
solution which is thermal AdS, is given by the region below line I. At any
finite temperature the parameters jump to values in regions (A) or (B).

The matrix model for the fixed charge is obtained by integrating
over the chemical potential. For this we need to know the exact
dependence of $a$ and $b$ on $\mu$ in this strong coupling regime
which we do not have. We note that the possibility (A) incorporates
a condition only on $a$. The condition on $b$ $(b>2(1-a))$ is always
satisfied in that range of $a$. This implies that we can
consistently assume $b$ to be independent of $\mu$. However the
possibility (B) implies that both $a$ and $b$ should be dependent on
$\mu$.

Equation(\ref{chpot}) shows the exact dependence of $a$ on $\mu$ in
the free theory. In this case, including only the charged scalars,
$a(\lambda,T,\mu)=[c(\lambda,T)+d(\lambda,T)\cos(\mu)]$. If we
assume this to hold in the strong coupling regime with $b$
independent of $\mu$ one can derive a model for the canonical (fixed
charge) ensemble. This will thus be consistent with (A). This fixed
charge model was studied in \cite{pb}. We will discuss the matrix
model corresponding to the Gauss-Bonnet black holes with fixed
charge in section (\ref{matrixch}). Before this in the next section we study the
model with fixed potential including $\alpha^{'}$ corrections.

%%%%%%%%%%%%%%%%%%%%%
%%%%%%%%%%%%%%%%%%%
%%%%%%%%%%%%%%%%%%%
\subsection{Including $\alpha^{'}$ corrections : Fixed potential}
\label{matrixpot}

As mentioned before, once we include $\alpha^{'}$ corrections along
with nonzero chemical potential, the coefficients in the
action (\ref{action}) depend also on $\alpha^\prime$. The equations of
motion are of the same form as given in the $\mu=0$ case:
\begin{eqnarray}
\rho F(\rho) &=&\rho  \quad, \quad
\quad\quad\quad 0 \leq \rho \leq \frac{1}{2},
\nonumber\\
&=& \frac{1}{4(1-\rho)}  \quad, \quad
 \frac{1}{2} \leq \rho \leq 1,
\label{eom}
\end{eqnarray}
where we have defined
\begin{equation}
F(\rho) = \frac{\partial S(\rho^2)}{\partial \rho^2} = N^2 [ 8 A_4
\rho^6  - 6 A_3 \rho^4 +  4A_2 \rho^2 + (1 - 2 A_1) ].
\label{polynomial}\end{equation} The potentials that follows from
the above action are given by \noindent \begin{eqnarray} V(\rho )
&=& - A_4 \rho^8 + A_3 \rho^6 - A_2 \rho^4 + A_1 \rho^2 \quad ,
\quad\quad 0 \leq
\rho \leq \frac{1}{2} \\
&=& - A_4 \rho^8 + A_3 \rho^6 - A_2 \rho^4  + (A_1 - 1/2) \rho^2 -
\frac{1}{4}\log[2(1-\rho)] + \frac{1}{8} \quad , \quad \frac{1}{2}
\leq \rho \leq 1 .\nonumber \label{potential}
\end{eqnarray}
As seen from (\ref{eom}) $\rho=0$ is always a solution. The action
(\ref{action}) evaluated at $\rho=0$ vanishes. For zero chemical
potential and $\bar\alpha$ this solution corresponds to the thermal
AdS on the bulk side\cite{aglw}.

In the analysis of phase structure we saw as the chemical potential
$\Phi$ and temperature $T$ vary we arrive different regions having
different thermodynamic features. Similarly in the matrix model as
chemical potential and temperature vary the coefficients of the
matrix model action vary as well. Thus if we consider a four
dimensional space corresponding to the four coefficients of
(\ref{action}) it is sectioned into various regions which are
analogous to the regions in the phase diagram. This is essentially
same as what we did graphically in Figure \ref{ab} for two
coefficients. Analogously, here we will identify different regions
in four dimensional space of the coefficients with various ranges of
temperature and chemical potential.

Let us begin with the various regions in the
($\Phi^2$-$\bar{\alpha}$) plane depending on the number and nature
of the solutions, as discussed in subsection (2.1) (see Fig.
\ref{alphaqphia}). In particular, there is a line in the
($\Phi^2$-$\bar{\alpha}$)-plane that separates region with three
solutions and that with one solution. Let us consider constraints
imposed on the coefficients $A_i$ that corresponds to these regions
or in other words, regions above and below the critical potential
$\mu_c$ at fixed $\lambda$. For that purpose it is useful to
consider the quadratic polynomial in $\rho^2$: $ f(\rho^2) =
(1/\rho)\frac{\partial}{\partial\rho} F(\rho)$. This is given by
$f(x) = 48 A_4 x^2 - 24 A_3 x + 8 A_2$. The zeroes of $f$ determine
the non-trivial turning points of $F$. We will see below that the
parameters at $T=0$ are continuously connected to the finite
temperature ones for $\mu <\mu_c$ but they are not so for $\mu
>\mu_c$ as we saw in absence of $\alpha^\prime$ correction.
\\

\begin{figure}[h]
\epsfxsize=16cm \centerline{\epsfbox{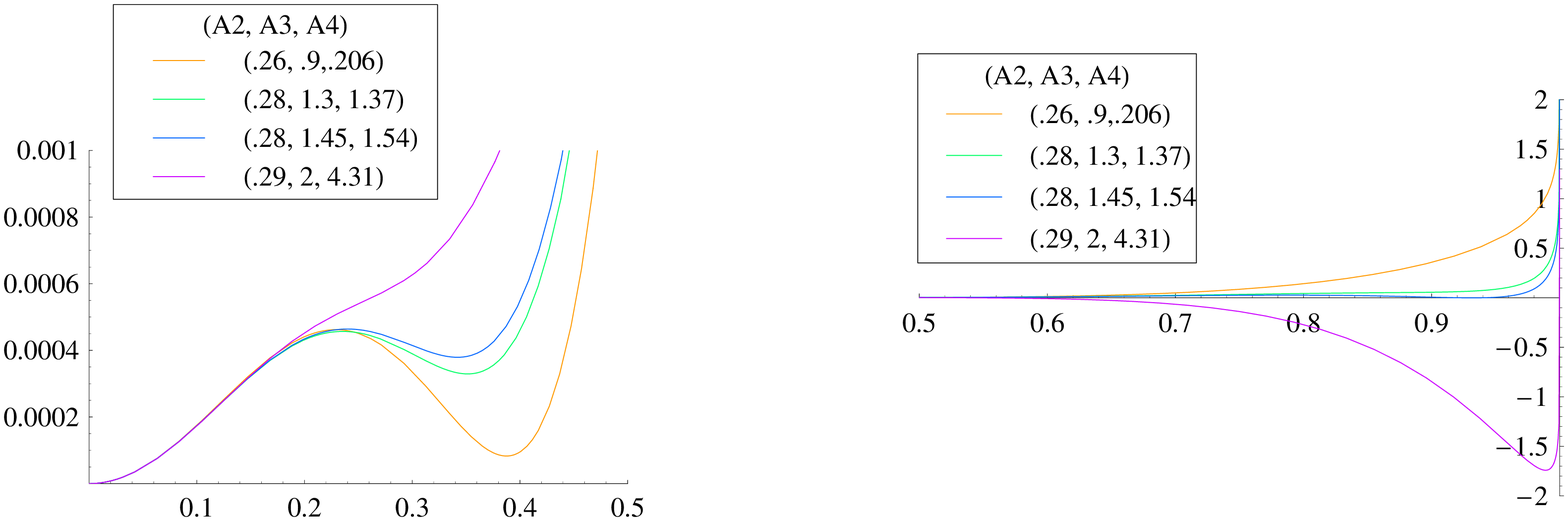}}
\caption{{\small{For $\mu < \mu_c,$ Potential as function of
$\rho$ for the ranges $0 \leq \rho \leq1/2$ and $1/2 \leq \rho \leq 1$.
The values of ($A_2$,
$A_3$, $A_4$ ) used in the plots are given above. $A_1=0.02$.}}}
\label{subcritical}\end{figure}

\noindent{\bf Sub-critical Potential $(\mu < \mu_c )$:} This
corresponds to the region I in Figure \ref{alphaqphia}. In this
range the behaviour is very similar to the case of zero chemical
potential.  As is evident from the plot in Figure \ref{phialpha}(c)
obtained on the gravity side we have two characteristic
temperatures, among others. One is $T =T_{N2}$ where the stable big
black hole and unstable intermediate black hole nucleates. Another
is $T=T_{N1}> T_{N2}$ where the intermediate unstable black hole
combines with stable small black hole and beyond that temperature
they cease to exist as solutions. This leads to three qualitatively
different ranges of temperature and in the following we discuss them
separately.\\

\noindent{\bf $T_{N2}< T < T_{N1}$:} Here on the gravity side we
have three black holes and so we expect at least three solutions of
the saddle point equation (\ref{eom}) obtained from the Matrix
model. The small black hole has a size of the order of
$\alpha^\prime$. In addition, as we have already seen in the
analysis of $\mu=0$ case\cite{dmms}, we have an (unstable) solution
which is a maximum of the potential. Since this solution does not
appear in the gravity that is beyond an analysis of order
$\alpha^\prime$. So we expect these two solutions occur generically
in the range $ 0 \leq \rho \leq \frac{1}{2} $. The other two
solutions which are analogue of unstable intermediate black hole and
stable big black hole, since they do have bulk analogue we expected
them to appear generically in the range $ \frac{1}{2} \leq \rho \leq
1 $. Therefore we get four solutions for this region.
So this region corresponds to the range of the coefficients $A_i$
for which (\ref{eom}) has four solutions within the domain of
$\rho$. From (\ref{eom}) we see this requires $f(\rho^2) = 0$ has
two positive solutions $0 \leq \rho_-^2 \leq \rho_+^2 \leq 1$ where
$\rho_-$ is a maximum and $\rho_+$ is a minimum of $F(\rho)$. That
$f(\rho^2)=0$ has two solutions requires
\begin{equation}
\triangle = 3 A_3^2 - 8 A_2 A_4 > 0 , \quad\quad A_4.A_2 > 0,
\label{4soln1}\end{equation} where the second condition is to ensure
that both the solutions $\rho_\pm^2$ are positive. That $\rho_\pm$
are minimum and maximum of $F$ respectively implies \quad
$F''(\rho_-) < 0$, $F''(\rho_+) > 0$ from which we obtain
\begin{equation} \rho_\pm^2 = \frac{A_3}{4A_4} \pm
\sqrt{(\frac{A_3}{4A_4})^2 - \frac{A_2}{6 A_4}}, \quad A_2
> 0. \label{4soln2}\end{equation}
Other conditions that we need to impose so that (\ref{eom}) admits
four solutions are $F(\rho_-) > 1$ and $F(\rho_+) < 1$, which when
written in terms of $A_i$'s become
\begin{eqnarray} \frac{A_2\cdot A_3}{3 A_4} > 2 A_1 + 16 A_4
\Big[(\frac{A_3}{4A_4})^2 - \frac{A_2}{6 A_4}\Big] \rho_-^2 \quad ,
\label{4soln3} \\
\frac{A_2\cdot A_3}{3 A_4} < 2 A_1 + 16 A_4
\Big[(\frac{A_3}{4A_4})^2 - \frac{A_2}{6 A_4}\Big] \rho_+^2 \quad
.\label{4soln4} \end{eqnarray} The above conditions
(\ref{4soln1},\ref{4soln2},\ref{4soln3}, \ref{4soln4}) ensure that
there exist a minimum ($\rho_{min}$) corresponding to the small
black hole and a maximum ($\rho_{max}$) that does not have a gravity
analogue. These solutions are in general different from $\rho_{\pm}$
and since $F$ is a cubic polynomial in $\rho^2$ the expressions are
complicated. In addition, on the gravity side we saw that stable
small black hole have positive energy. So we expect the solution of
the saddle point equation that correspond to this stable small black
hole should have positive energy. So we need at the minimum
\begin{equation}
V_1(\rho_{min}) > 0 \quad . \label{positiveenergy}\end{equation}
whose explicit form is not very illuminating.

In addition as we said from gravity analysis we expect two solutions
should appear, generically, in the first half of the domain of
$\rho$ while two other in the second half. We consider two halves of
the domain of $\rho$ separately. Let us begin with the range $0 \leq
\rho \leq \frac{1}{2}$. That we have two solutions of the saddle
point equation (\ref{eom}) in this range imposes two further
conditions:
\begin{equation}  \rho_- < \frac{1}{2}, \quad F(\frac{1}{2}) < 1.
\end{equation} ( In this form it is easier to obtain the restriction
on the parameters. Otherwise we could write $ 0 < \rho_{min} ,
\rho_{max} < \frac{1}{2}$ which are more direct but the expressions are
cumbersome.) These conditions, when written in terms of the
parameters, reduce to the following two equations:
\begin{eqnarray}
\frac{A_3}{4A_4} -
\sqrt{(\frac{A_3}{4A_4})^2 - \frac{A_2}{6 A_4}} & < & \frac{1}{4}\quad ,\label{reg1a} \\
2 A_1 + \frac{3 A_3}{8} & > & A_2 + \frac{A_4}{8}.
\label{reg1b}
\end{eqnarray}
These conditions
(\ref{4soln1},\ref{4soln2},\ref{4soln3},\ref{4soln4},\ref{positiveenergy})
are real inequalities in four parameters. Each of them will give
rise to one (or more) codimension 1 wall(s) in the four dimensional
parameter space described by ${A_1, A_2, A_3, A_4}$. Since this
involves four parameters it is difficult to have a graphical
representation of it. As we see each of the inequalities corresponds
to wall(s) which we cross when we go beyond this range of
temperature and chemical potential. However, crossing the wall
corresponding to (\ref{positiveenergy}) will take us to some
unphysical region as that implies the stable small black hole has
energy less than that of AdS which implies on the gauge theory side
 absence of confinement in low temperature. The
equations (\ref{reg1a} and \ref{reg1b}) also give rise to codimension one
walls but a more elaborate analysis is required to understand the
correct significance of them.

We expect the other two solutions to appear in the other half namely
$\frac{1}{2} \leq \rho \leq
 1 $. On this part the situation is pretty much similar to that of
 $(a,b)$-model. This requires we have one solution $\frac{1}{2} \leq \rho_1 \leq 1$
 such that it satisfies
 \begin{equation}
 F(\rho_1) = \frac{1}{4 \rho_1 ( 1 - \rho_1)}
 \quad , \quad\quad
F'(\rho_1) > \frac{1}{4}(- \frac{1}{\rho_1^2} + \frac{1}{( 1 -
\rho_1)^2} ), \label{4soln5}\end{equation} where this $\rho_1$ will
appear as the analogue of the unstable intermediate black hole
solution that we got on the gravity side. Actually, this condition
is sufficient to ensure that there is the analogue of stable big
black hole too. This condition (\ref{4soln5}) along with
(\ref{4soln1},\ref{4soln2},\ref{4soln3},\ref{4soln4}) completes the
list of necessary and sufficient condition to have four solutions of
the saddle point equations. As temperature increases, $\rho_1$ will
decrease. At Gross-Witten temperature
\cite{Gross:1980he,Wadia:1980cp} $T=T_g$ this reaches the
lower boundary of this region $\rho = \frac{1}{2}$. At $T=T_g$ the
restriction on the parameters can be written as
\begin{equation}
 2 A_1 + \frac{3A_3}{8} = A_2 + \frac{A_4}{8}
 \quad , \quad
 3 A_4 + 8 A_2 > 6 A_3 .
 \end{equation}
 This corresponds to a third order phase transition.
 \\

\noindent{\bf $ T \leq T_{N2}$:} In this range of temperature the
conditions that we obtain on the first half of the domain of $\rho$
remain the same. On the second half of the domain the two solutions
will merge at $T=T_{N2}$. This implies in the four parameter space
we are on the wall that corresponds to (\ref{4soln5}). The following
equations corresponds to $T_{N2}$:
\begin{equation}
F(\rho) = \frac{1}{4 \rho ( 1-\rho)} \quad ,\quad F'(\rho_1) =
\frac{1}{4}(- \frac{1}{\rho_1^2} + \frac{1}{( 1 - \rho_1)^2} ).
\end{equation}
As temperature decreases the value of $F(\rho)$ will monotonically
decreases and there will be no solution in the second half.
\\

\noindent {\bf $ T \geq T_{N1}$:} Here we have no solution in the
first half of domain of $\rho$ at $T > T_{N1}$. But that can happen
in two possible ways and we discuss them in the following. One
possibility is the two extrema of $V_1(\rho)$ merge at $T=T_{N1}$
which implies we cross the wall corresponds to (\ref{4soln3}). In
terms of the parameters this means at $T=T_{N1}$
\begin{equation}
\frac{A_2 A_3}{3 A_4} = 2 A_1 + 16 A_4 \Big[{(\frac{A_3}{4A_4})^2 -
\frac{A_2}{6 A_4}}\Big] \rho_-^2 .
\end{equation}
This is a consistent possibility that can occur in the matrix model.
Another significance of this equation is this corresponds to the
bound beyond which we get the features of $(a,b)$-model and so this
region sits in the same universality class of that of $(a,b)$-model.

The second possibility is crossing the wall corresponding to
(\ref{4soln4}), where the minimum of $V_1$ meets the maximum of
$V_2$. However,  In terms of the coefficients, then, at $T_{N1}$ we
have \begin{equation} \frac{A_2 A_3}{3 A_4} = 2 A_1 + 16 A_4 \Big[
{(\frac{A_3}{4A_4})^2 - \frac{A_2}{6 A_4}}\Big] \rho_+^2 .
\end{equation}
This possibility seems more plausible as it matches with what we saw
on the gravity side. There at $T = T_{N1}$ the unstable intermediate
black hole merges with the stable small black hole. At this point we
should comment on the relative values of $T_{N1}$ and $T_g$. If
$T_{N1} < T_g$ this merging occurs before the system reaches it
Gross-Witten point and therefore there will be no Gross-Witten
transition. On the other hand for $T_{N1}
> T_g$ the intermediate solution has already undergone
the GW transition and becomes a black hole of the order of
$\alpha^\prime$. The other possibility is if $T_{N1}=T_g$ where the
small black hole merges with the intermediate black hole exactly at
the Gross-Witten point or the two walls corresponding to
(\ref{4soln4}) and (\ref{reg1a}) merge.
\\

\begin{figure}[h]
\epsfxsize=16cm \centerline{\epsfbox{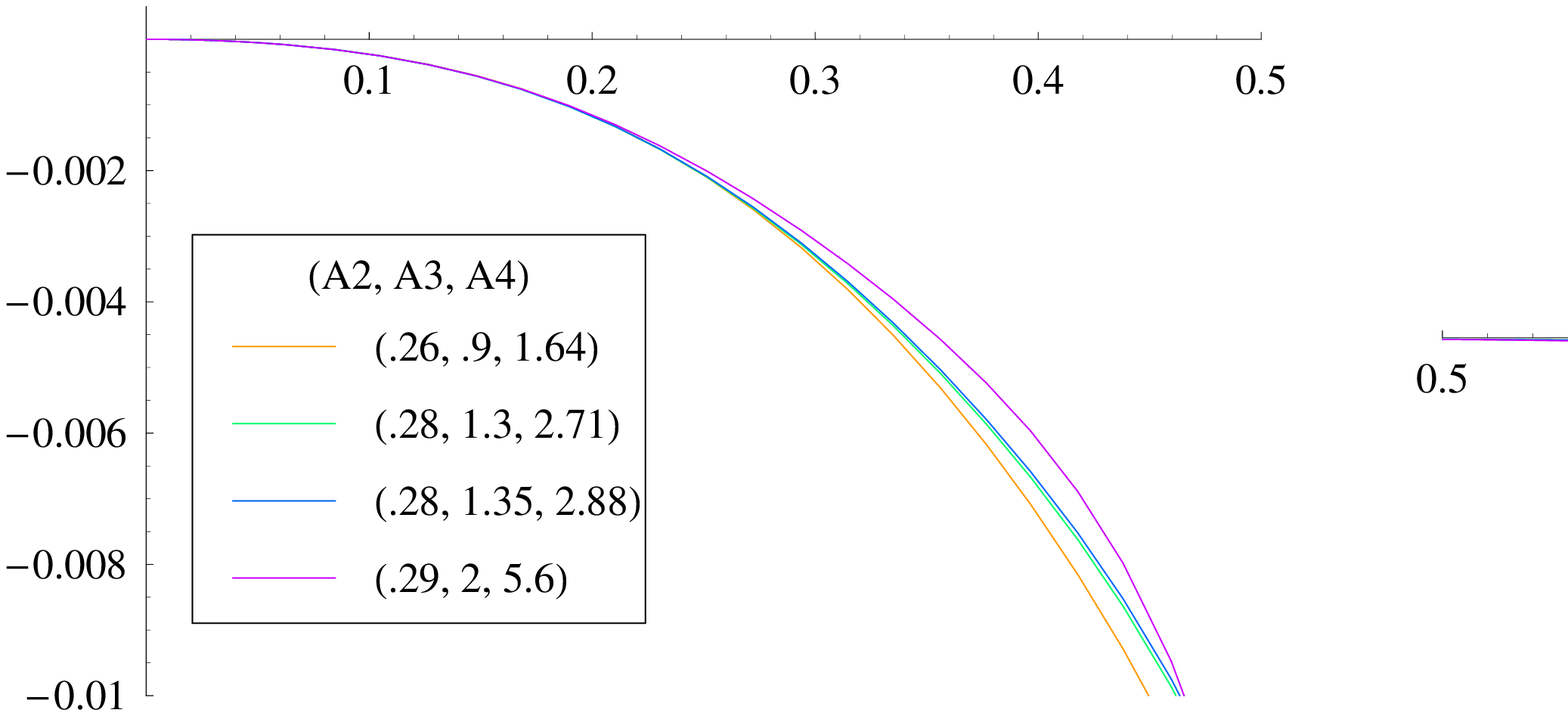}}
\caption{{\small{For $\mu > \mu_c,  A_1<0 , \triangle<0$: Potential as function of
$\rho$ for the range $0\leq\rho \leq 1/2 $ and $1/2 \leq\rho \leq 1 $ .The values of ($A_2$, $A_3$, $A_4$ ) used in the plots are given above. $A_1=-0.02$.}}}
\label{supercriticala1}\end{figure}

\noindent{\bf Super-critical Potential $(\mu > \mu_c)$:} This region
corresponds to the parameter space above region I in Figure
\ref{alphaqphia} and therefore should have a single minimum of the
potential. In the matrix model like the $\mu < \mu_c$ region here
also we have an additional unstable maximum of the potential that
does not have any bulk analogue and so it is beyond the gravity
analysis of the $\alpha^\prime$ order. Once again we restrict this
in the region, $\rho<1/2$ so that this corresponds to some stringy
phase as suggested in \cite{aglw}. Therefore, similar to the
analysis for $\alpha^\prime = 0$ ( and unlike the $\mu < \mu_c$ case
) we consider two possible scenarios depending on whether the
unstable maximum lies at $\rho=0$ or it lies away from $\rho=0$. In
the following we discuss the two scenarios separately.
\\

\noindent {\bf (A) Unstable maximum is at $\rho=0$ $(A_1 < 0)$:} If
we have $(A_1 < 0)$ the unstable maximum will always be at $\rho=0$.
In addition, in the matrix model we expect to have one and only one
stable minimum for $\rho > 0$ (The energy of this black hole could
be greater or less than that of thermal AdS depending on the values
of $\mu$ and $\lambda$ (see Figure \ref{ftphi}). The condition $A_1
< 0$ alone does not ensure this and we need to impose additional
constraints on $A_2$, $A_3$ and $A_4$. We note that there are two
ways in which one can ensure that there is only one stable black
hole solution. We discuss them separately in the following:

One possibility is, at $\mu=\mu_c$ the three solutions merge into
one at some value of $\rho = \rho_+ > 0$. A similar merging occurs
on the gravity side when the chemical potential reaches its critical
value. Since the corresponding solution is manifested in the gravity
limit we expect that $\rho_+ > 1/2$. Then, at $\mu=\mu_c$, apart
from the condition $A_1=0$, $A_2$, $A_3$ and $A_4$ satisfy,
 \beqal{merge}
{V^{'}}(\rho_+)=0\mbox{\hspace{0.1in};\hspace{0.1in}}
 {V}^{''}(\rho_+)=0\mbox{\hspace{0.1in};\hspace{0.1in}}
{V}^{'''}(\rho_+)=0 \mbox{\hspace{0.1in}for\hspace{0.1in}} \f{1}{2}
\le \rho_+ \le 1. \eeqa Thus we obtain the parametric solutions of
$A_2(\rho_+)$, $A_3(\rho_+)$ and $A_4(\rho_+)$ from (\ref{merge})
which corresponds to the line separating the number of solutions in
gravity in Figure \ref{alphaqphia}. We do not write these
parametric equations here as they are not compact enough. For,
$\rho_+=0.8$ we have, $A_1= 0$, $A_2=0.583$, $A_3=1.481$,
$A_4=1.293$.

Examining the equation of motion (\ref{eom}), one can find an
indirect way to obtain a (slightly more general) constraint in
compact form. This ensures there is exactly one solution at some
point $\rho
> 0$ in addition to the one at $\rho=0$. However, that the three
solutions merge at this point is not guaranteed. If we relax the
first inequality of (\ref{4soln1}) by making the discriminant
$\triangle$ negative so that there is no turning point of $F$ then we
are left with only a single solution. This can be thought of as
crossing the wall corresponds to the first condition of
(\ref{4soln1}) with $\mu=\mu_c$. The merging of the three solutions
at this point appears as a special case of it and so that condition
is more ramified. We have plotted the associated potentials in
Figure \ref{supercriticala1}.

\begin{figure}[h]
\epsfxsize=16cm \centerline{\epsfbox{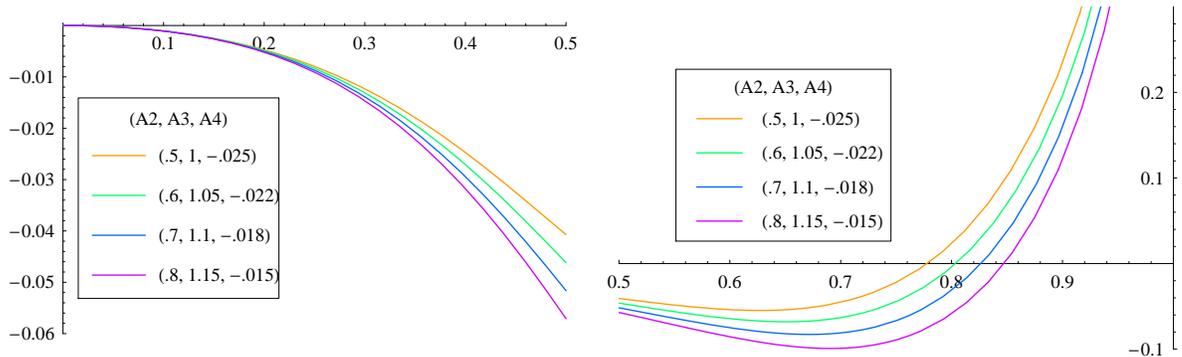}}
\caption{{\small{For $\mu > \mu_c, A_1<0, A_4<0$: Potential as
function of $\rho$ for the range $0\leq\rho \leq 1/2 $. The values
of ($A_2$, $A_3$, $A_4$ ) used in the plots are given above.
$A_1=-0.1$.}}} \label{supercriticala2}\end{figure}

The other possibility that one can consider to obtain a single
minimum is setting $A_4(\mu=\mu_c)=0$ and $A_4(\mu
> \mu_c) < 0$. This amounts to relaxing the second inequality of
(\ref{4soln1}) so that one of the turning point becomes imaginary.
This corresponds to identifying $\mu=\mu_c$ with the second
condition of (\ref{4soln1}). For this choice, as $\mu$ increases
beyond its critical value, one of the turning point comes down to
$\rho=0$ and then disappears. This possibility is not continuously
connected with the above mentioned constraint, which gives rise to
merging of three solutions. However, this possibility can give a
simple form for the fixed charge case. We have plotted the related
potentials in Figure \ref{supercriticala2}.
\\

%\begin{figure}[h]
%\epsfxsize=16cm \centerline{\epsfbox{supercriticala2.eps}}
%\epsfxsize=16cm \centerline{\epsfbox{potentialcqq2.eps}}
%\psfig{Potential03.epsi,width=6in}
%\caption{{\small{For $\mu > \mu_c, A_1<0, A_4<0$: Potential as
%function of $\rho$ for the range $1/2 \leq\rho \leq 1 $. The values
%of ($A_2$, $A_3$, $A_4$ ) used in the plots are given above.
%$A_1=-0.1$}}} \label{supercriticala2a}\end{figure}

 \noindent {\bf (B) Unstable
maximum lies at $ 0 < \rho < 1/2$ $(A_1>0)$}: In this scenario for
$\mu
> \mu_c$ the unstable maximum is in the range $0 < \rho < 1/2$.
The usual saddle point $\rho=0$ is also there but this time it
corresponds to the thermal AdS. In addition, there exists only one
minimum. (The energy of this minimum is either greater or less than
that of AdS at low temperatures. If the energy is greater than that
of AdS energy, at high temperature, the black hole undergoes Hawking
Page transition.) In order to make sure that this is the one and the
only one minimum, once again we identify $\mu_c$ to be the limit
where three saddle points merge. However unlike the previous case
where the small maximum was at $\rho=0$, here it lies in $0 < \rho <
1/2$. This configuration satisfies,
 \beqal{merge1}
{V^{'}}(\rho_+)=0\mbox{\hspace{0.1in};\hspace{0.1in}}
 {V}^{''}(\rho_+)=0\mbox{\hspace{0.1in};\hspace{0.1in}}
{V}^{'''}(\rho_+)=0 \mbox{\hspace{0.1in}for\hspace{0.1in}}
\f{1}{2} \le \rho_+ \le 1.\non
\mbox{and \hspace{0.1in}} {V^{'}}(\rho_-)=0
\mbox{\hspace{0.1in}for\hspace{0.1in}}
0 \le \rho_- \le \f{1}{2}.
\eeqa

\begin{figure}[h]
\epsfxsize=16cm \centerline{\epsfbox{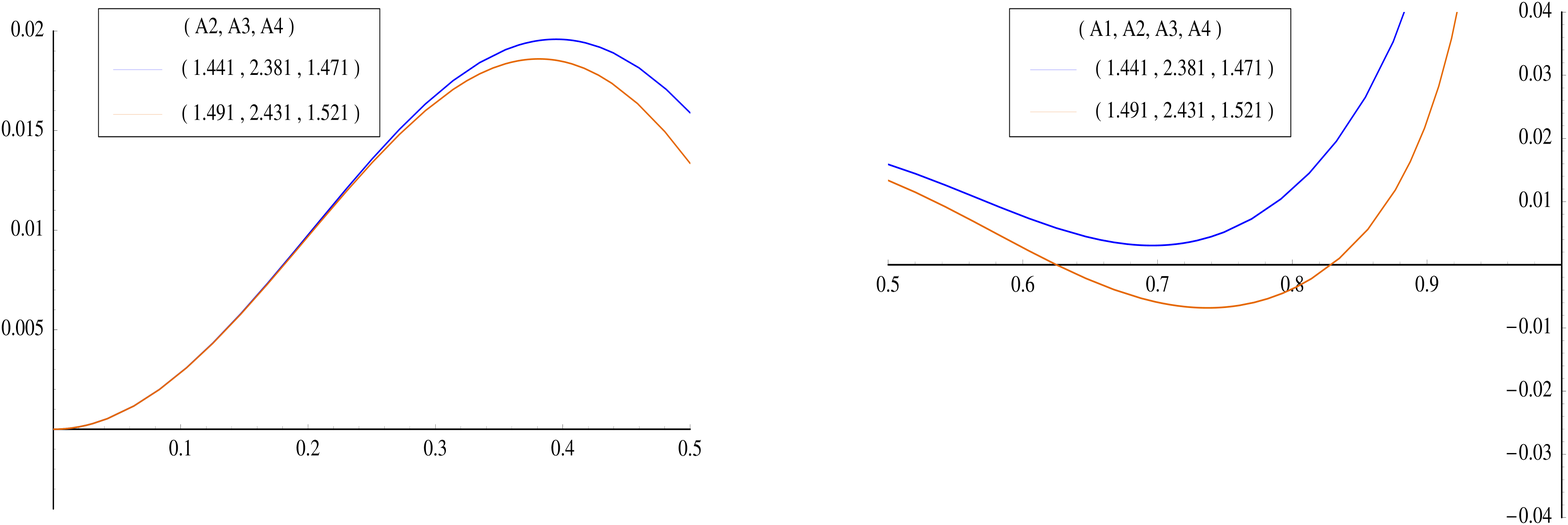}}
\caption{{\small{For $\mu > \mu_c$ and $A_1>0$: Potential as function of
$\rho$ for the range $0\leq\rho \leq 1/2 $ and $1/2 \leq\rho \leq 1$.
The values of ($A_2$,
$A_3$, $A_4$ ) used in the plots are given above. $A_1=0.298$.}}}
\label{yhp}\end{figure}

Apart from being functions of $\rho_+$ the $A_i$'s are now also
functions of $\rho_-$. With $\rho_-=0.5$ and $\rho_+=0.8$ we get,
$A_1=0.418$, $A_2=1.561$, $A_3=2.501$ and $A_4=1.691$. In this case
the energy of the merging point ($\rho_+$) is positive. This gives
rise to a stable minimum with positive energy as we move beyond
$\mu_c$. We now vary the parameters with increasing temperature and
see that the minimum crosses zero corresponding to the Hawking-Page
transition in the bulk. The plots for $\mu > \mu_c$ are shown in
Figure \ref{yhp}.

\begin{figure}[h]
\epsfxsize=16cm \centerline{\epsfbox{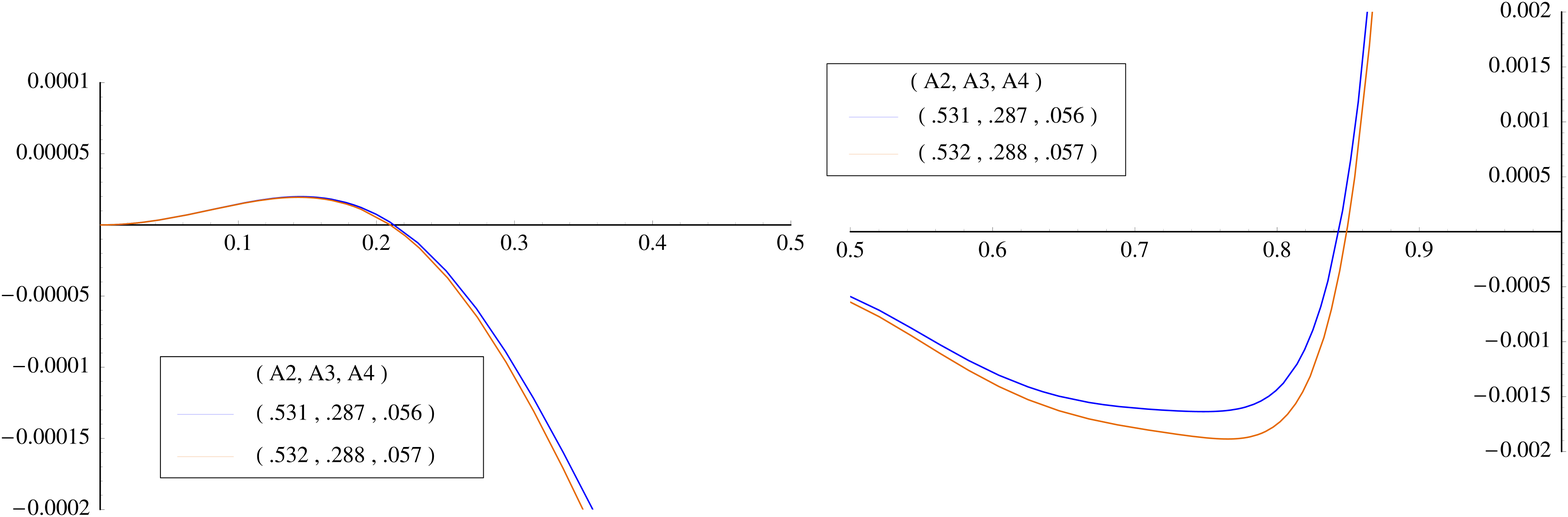}}
\caption{{\small{For $\phi > \phi_c, A_1>0$: Potential as function of
$\rho$ for the range $0\leq\rho \leq 1/2 $ and $1/2 \leq\rho \leq 1$.
The values of ($A_2$,
$A_3$, $A_4$ ) used in the plots are given above. $A_1=0.002$.}}}
\label{nhp}\end{figure}

One can similarly derive the parameters for which the energy of the
merging point ($\rho_+$) is less than zero. This is given by,
$\rho_-=0.4$ and $\rho_+=0.72$ which gives $A_1=0.006$, $A_2=0.06$,
$A_3=0.291$ and $A_4=0.535$. Note that for $\rho_-=0$ we get
$(A_1=0)$, which is the possibility (A). As we move away from the
critical potential, this gives rise to a minimum with energy less
than AdS. With further variation of the parameters this saddle point
goes deeper as shown in Figure \ref{nhp}. This can be mapped to
the variation of the minimum with temperature as it happens for the
black hole in gravity.

In the above discussion we consider, on the matrix model side, the
various possibilities, which can reproduce the bulk behaviour that we
obtained in the analysis on the gravity side. The different
possibilities gives rise to different ranges of the parameters, of
which some are mutually exclusive. With this amount of input it is
not possible to decide which one is the correct behaviour. We need
the explicit dependence of $A_i$'s on $\mu$ in the strong coupling
regime. Perhaps a weak coupling calculation of the associated terms
in the model can serve as a clue.

\subsection{Including $\alpha^{'}$ corrections : Fixed charge}\label{matrixch}

We have seen in the last section that when restricted to various values of
the parameters $A_i$, the matrix model (\ref{action}) reproduces the features
of gravity below and above the critical potential, $\mu_c$. In general thus
$A_i$'s will depend on the chemical potential. In the study of the matrix model
for fixed charge, the explicit knowledge of the dependence of $A_i$'s are
necessary. Out of the cases studied in the earlier section, there is one that
allows us to construct a toy model for the fixed charge consistently. This
is given by the second possibility of (A), when both $A_1$ and $A_4$ become
negative above the critical potential.

In the following, we will assume the result from the zero coupling to be valid
in this strong coupling regime. In the free theory, $A_1$ is given by,
$(1-2A_1) = [c+d\cos(\mu)]$ when matter fields consist of only charged scalars.
Since we are only interested in the qualitative features of this model, for
simplicity we will take, $A_4=a_4(\lambda,T)\cos(\mu)$ with
$A_2$ and $A_3$ functions only of $\lambda$ and $T$. The action with this
dependence on the potential is,

\begin{equation}
S(\rho^2)= 2 N^2 \left[a_4 \cos(\mu)\rho^8  -  A_3 \rho^6 +  A_2 \rho^4 +
\left(\frac{c+d\cos(\mu)}{2}\right) \rho^2 \right]. \quad
\label{actionpot}
\end{equation}

The partition function for the canonical ensemble is obtained by
integrating over the chemical potential. Hence following \cite{pb}, we write,

\beqa
Z(Q,T,\lambda)= \int d\mu \exp(-i\mu Q)\int d\rho \exp\{-N^2 V(\rho,\mu)\},
\eeqa

\noindent where, the potential is,

\beqal{chargepot}
V(\rho)&=&-\f{1}{2N^2}S_{eff}(\rho^2)+\f{1}{2}\rho^2
\quad, \quad
\quad\quad\quad 0 \leq \rho \leq \frac{1}{2},
\nonumber\\
&=&-\f{1}{2 N^2}S_{eff}(\rho^2)-
\frac{1}{4}\log[2(1-\rho)] + \frac{1}{8} \quad , \quad \frac{1}{2}
\leq \rho \leq 1,
\eeqa

\noindent with,

\beqa\label{chargeeff1}
S_{eff}(\rho^2)&=&  N^2\left[c\rho^2 +2 A_2 \rho^4 -2 A_3 \rho^6\right] +
\log\left[I_Q\left(N^2(2a_4 \rho^8 +d\rho^2)\right)\right].
\eeqa

\noindent Here $I_Q(x)$ is a Bessel Function.  
We are interested in the the large $N$ limit. This is obtained by 
keeping in mind that $Q^2\sim {\cal O}(N^2)$
so that $Q^2=N^2q$ where $q\sim{\cal O}(1)$. The resulting effective action
in the large $N$ limit is thus,

\begin{eqnarray}\label{chargeeff}
S_{eff}(\rho^2)&=&  N^2\left[c\rho^2 +2 A_2 \rho^4 -2 A_3 \rho^6\right] +
\\\nonumber
&+& N^2 q
\left[\left(1 + \frac{ \rho^4 (d+ 2 a_4 \rho^6)^2 }{q^2}\right)^{\frac{1}{2}}
+ \log\left[\frac{\frac{ \rho^2 (d+ 2 a_4 \rho^6)}{q}}
{ 1 + \{1+\frac{ \rho^4 (d+ 2 a_4 \rho^6)^2 }{q^2}\}^{\frac{1}{2}}}\right]
\right]\\\nonumber
&+& {\cal O}\left(\f{1}{N^2}\right).
\end{eqnarray}

\noindent Defining, $F(\rho)= \pa{S(\rho^2)}/\pa{\rho^2}$, the equation of motion
is given by (\ref{eom}).

\begin{figure}[h]
\epsfxsize=10cm \centerline{\epsfbox{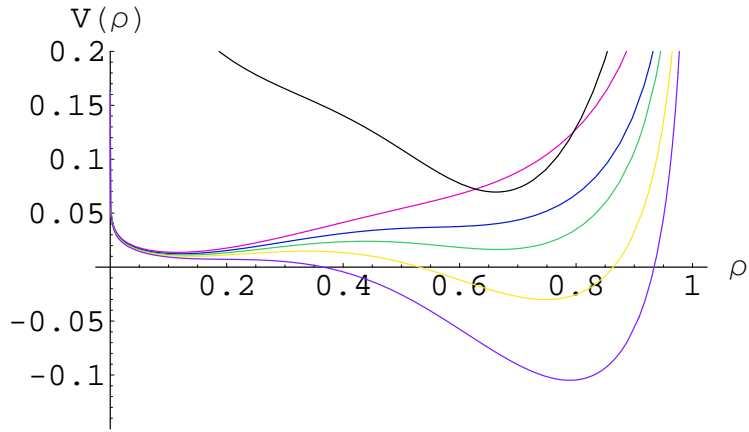}}
\caption{Coloured curves: Potential for various temperatures 
corresponding to the
variations of $A_1$, $A_2$, $A_3$, $a$, $c$ and $a_4$ for $q <q_c$
Black curve: Potential for $q >q_c$.}
\label{RNfchmatrix}\end{figure}

The phase structure is qualitatively the same as when there are no
$\alpha^{'}$ corrections. Unlike the fixed potential thermal AdS is
not a solution. It is easily seen from (\ref{chargeeff}) that
$\rho=0$ is not a solution of the equation of motion. We have seen
in section \ref{can}, that there exists a critical charge for a
fixed $\bar{\alpha}$ beyond which there is only one solution. Below
this critical charge there are a maximum of three solutions. See
equation (\ref{qbaralphab}) and Figure \ref{alphaqphib}. 
In the matrix model, this critical charge ($q_c$) is given by the merging of 
the three saddle points of (\ref{chargepot}).

Figure \ref{RNfchmatrix} shows the thermal history for the canonical 
ensemble. The coloured curves are for $q < q_c $ for some fixed $\lambda$.  
This corresponds to region $I$ in Figure \ref{alphaqphib}.
At low temperatures, there is a single minimum in the range $0<\rho<1/2$
(curves in pink). This
corresponds to the small black hole. Two new saddle points appear at $T=T_3$
with the new minimum in the range $ 1/2<\rho<1$ (shown in blue)
. These correspond to
the intermediate (unstable) and large black holes. The small black hole has
lower energy upto  $T_c$ (the green curve), beyond which the large black hole 
phase is favoured (yellow).
Thus there is a phase transition at $T_c$ corresponding to the one in the bulk.
The small and the intermediate black holes disappear at $T_1$ and at high
temperature, the large black hole (the phase with the saddle point in the range
 $1/2<\rho<1$) remains (shown in purple).

The black curve in Figure \ref{RNfchmatrix} shows the single saddle point
when $q > q_c $. This corresponds to the single large black hole that exists
above the critical charge for fixed $\bar{\alpha}$, in the region $II$ of 
Figure \ref{alphaqphib}.

\end{document}